\documentclass[%
superscriptaddress,
 amsmath,amssymb,
aps,
floatfix,
]{revtex4-2}
\usepackage{filecontents}
\usepackage{graphicx}
\usepackage{dcolumn}
\usepackage{bm}
\usepackage{subfigure}
\usepackage{subfloat}
\usepackage{overpic}
\usepackage{tikz}
\usepackage{color}
\usepackage{amsmath}
\usepackage{amssymb}
\usepackage[normalem]{ulem}

\newcommand\h{$^1\text{H}$}

\usepackage{xr}

\makeatletter
\newcommand*{\addFileDependency}[1]{
\typeout{(#1)}
\@addtofilelist{#1}
\IfFileExists{#1}{}{\typeout{No file #1.}}
}\makeatother
\newcommand*{\myexternaldocument}[1]{%
\externaldocument{#1}%
\addFileDependency{#1.tex}%
\addFileDependency{#1.aux}%
}
\myexternaldocument{letter_SI}
\usepackage[normalem]{ulem}

\begin{document}

\preprint{APS/123-QED}

\title{ Multi-mode masers of thermally polarized nuclear spins in solution NMR}

\author{Vineeth Francis Thalakottoor Jose Chacko}
\email{vineeth.thalakottoor@ens.psl.eu}
\affiliation{%
 Laboratoire des Biomolécules, LBM, Département de Chimie, Ecole Normale Supérieure, PSL University, Sorbonne Université, CNRS, 75005 Paris, France
}%
\author{Alain Louis-Joseph}
\email{alain.louis-joseph@polytechnique.edu}
\affiliation{
 Laboratoire de Physique de la Matière Condensée, UMR 7643, CNRS, École Polytechnique , IPP 91120 Palaiseau, France
}%
\author{Daniel Abergel}%
\email{daniel.abergel@ens.psl.eu}
\affiliation{%
 Laboratoire des Biomolécules, LBM, Département de Chimie, Ecole Normale Supérieure, PSL University, Sorbonne Université, CNRS, 75005 Paris, France
}%
\date{\today}
\begin{abstract}
We present experimental   single and multimode  sustained \h\;NMR masers  in solution on thermally polarized spins at  room temperature and 9.4 T achieved through the electronic control of radiation feedback (radiation damping). Our observations illustrate the breakdown of the usual three-dimensional Maxwell-Bloch equations for radiation feedback and a simple toy model of few coupled classical moments is used to interpret these experiments.
\end{abstract}

\maketitle

Nonlinearities in nuclear magnetic resonance spectroscopy have attracted much attention in the past decades.
Such non conventional behaviours have their origins in cooperative phenomena that involve either the coupling of the spins with the detection circuit - the resonating cavity- in the case of the so-called "radiation damping",\cite{bloembergen_RD_1954}  or distant dipolar field effects involving the whole sample\cite{Deville_1979} - a nonlocal phenomenon -, or a combination of both.
In high-resolution solution NMR, radiation damping (RD) gives rise to the typical maser induction decay\cite{bloom57,szoke_1959}, as well as other  unusual observations,\cite{mccoy_1990, abergel_1992, abergel_1994,barjat_RDeffect_1995, mao_RDeffects_1994, Ball_RDeffect_1996,Pelupessy_RDeffect_2022} and different techniques have been developed to suppress it by using gradient, Q-switching or electronic feedback.\cite{Sklenar_RDsuppress_1995, anklin_RDsuppress_1995,Broekaert_RDsuppress_1995,Louis-Joseph_RDsuppress_1995}
Radiation feedback can in some cases be associated to distant dipolar field effects to yield even more complex phenomena\cite{lin_2000}.
Sustained NMR masers have also be observed in hyperpolarization experiments in solution, such as $^{129}$Xe spin exchange optical pumping experiments, SABRE experiments in low field\cite{suefke_2017}, as well as in the context of solid-state dynamic nuclear polarization at liquid helium\cite{Bosiger1977,Weber2019} or in magic-angle spinning experiments\cite{hope2021}.
The interplay of radiation damping and the distant dipolar field have also been pinpointed in this context.\cite{thalakottoor_dipole_2023}.

The existence of sustained maser pulses caused by radiation damping in these experiments is quite remarkable and reflects the coexistence of two antagonist phenomena that drive the spins to opposite polarization states, therefore  the magnetization, towards a direction that is either aligned or opposite to the magnetic field.
Indeed, the radiofrequency back action field from the cavity, slaved to the magnetization, brings the latter to the north pole of the Bloch sphere, a stable direction.\cite{bloom57}
An additional ingredient is therefore required to reestablish the conditions that  lead to a maser pulse.
In the context of DNP, such a process is provided by the hyperpolarization, when the microwave source constantly polarizes the nuclear spins "negatively", that is, opposite to the equilibrium direction, or when, once polarized, additional spin  species ($^2$H) acts as a polarization reservoir for the spin species coupled to the probe ($^1$H).\cite{Bosiger1977,Weber2019}
Hyperpolarizaton processes leading to NMR masers involve the combination of a nonlinear dynamical process and a repolarization one.\cite{suefke_2017}
Alternatively, it has been shown that in the context of solution state NMR for thermally polarized water, that a set of antagonist phenomena could be achieved by $T_1$ relaxation and an electronically generated feedback field from the probe, that instead of flipping the magnetization to the north pole of the Bloch sphere, brings it to the south pole.\cite{abergel_maser_2002}

In this letter, we report the observation of a multi-mode sustained maser in solution NMR at room temperature and generated by an electronic manipulation of the cavity signal.\cite{Louis-Joseph_RDsuppress_1995}
We show the surprising result that for partially resolved lines, such as due to $B_0$ inhomogeneities, sustained masers with unattenuated steady state are observed, which can be proven by a mathematical analysis to be in contradiction with the usual Maxwell-Bloch equations.
Instead, one must instead  consider explicitly the collective effect of the couplings of several moments with a distribution of Larmor frequencies to the probe, an aspect previously suggested by numerical simulations.\cite{abergel_maser_2002,thalakottoor_dipole_2023}
Moreover, a simple toy model of two moments undergoing the same common feedback field from the probe is shown to account for our observations and illustrates the transition between the conventional Maxwell-Bloch behaviour and the actual limit cycle steady state.

Radiation damping (RD)  results from the strong coupling between the precessing magnetization and the high-quality factor (Q) resonant LC circuit of the detection circuit.
The current induced by a large magnetization in the detection coil results in a magnetization-dependent feedback field perpendicular to the transverse component.
The dynamics of the magnetization subject to radiation damping  is modeled by the well-known Bloch-Maxwell equations.\cite{bloembergen_RD_1954, bloom57,Vlassenbroek_RD_1995} 
These describe the dynamics of a large magnetization ${\bf m}$ strongly coupled to a high-Q resonant circuit, and  acted upon by a radiation feedback field ${\bf B}_{\textsc{fb}} = G m(t)  e^{-i \psi}$, where $m(t) = \langle m_x\rangle+ i \langle m_y\rangle$  is the transverse magnetization, and characterized by its gain $G =   \frac{\mu_{o} \eta Q}{2}$ and phase   $\psi=-\pi/2$, when the NMR detection circuit is tuned to the spins Larmor frequency.
It is also possible to generate electronically an "artificial" feedback field in order to modulate (either suppress or enhance) the RD field by controlling $G$ and $\psi$.\cite{Louis-Joseph_RDsuppress_1995}
Thus, in the presence of an applied rotating radiofrequency field ${\bf B_{1}}=B_1 e^{i(\omega t +\psi_1)}$ with angular frequency $\omega$, the Bloch-Maxwell equation in the rotating frame can be written as:
\begin{eqnarray}
\nonumber \frac{d}{dt} {\bf m}(t) &=& \gamma {\bf m}(t) \times \left [ -(\delta\omega/\gamma) \hat{\bf z}+ {\bf B}_1 +{\bf B}_{\textsc{fb}} \right ]\\
&-& \gamma_2 m(t) -\gamma_1 \left [ m_z(t)-  m_{o}  \right ] \hat{\bf z}  
 \label{eq:bloch-maxwell-r}
\end{eqnarray}
where the offset  $\delta\omega = \omega_{o} - \omega$, $\gamma$ is the gyromagnetic ratio, $\gamma_1$ and $\gamma_2$ are longitudinal and transverse relaxation rates. 
%
The Maxwell-Bloch equations can be solved only in particular cases\cite{bloom57,Vlassenbroek_RD_1995, barbara_1992} but in general require numerical investigations.
They are nevertheless amenable to a qualitative analysis,\cite{abergel_jcp_2000, abergel_maser_2002, Weber2019}
which predicts a sustained maser evolution that eventually leads to a constant transverse amplitude of the magnetization, provided that $m_0 \times \sin \psi  > 0$. This condition implies that two antagonist processes, i.e.,  radiation feedback (the nonlinear process) and longitudinal relaxation (the dissipative process), coexist.
In conditions prevailing in solution state NMR, where the spins relax to thermal equilibrium ($m_0>0$), requires that  the feedback field drives the magnetization towards the $-z$ direction ($\sin \psi  > 0$).
Alternatively, this necessary condition is also met in hyperpolarization experiments, when the nuclear spins are polarized negatively ($m_0<0$) and the radiation damping field brings the magnetization towards the $+z$ direction ($\sin \psi  < 0$).
This behaviour was observed in low temperature DNP,\cite{Bosiger1977,Weber2019, thalakottoor_dipole_2023} or SABRE\cite{appelt_jmr_2021} experiments.

When several resolved resonance lines are present,  each ${\bf m}_k$ contributes to the total feedback field $ {\bf B}_\textsc{fb} = G e^{- i \psi}\sum_k m_k(t)$, where $m_k(t)$ is the the transverse magnetization of spin $k$.
The associated moments ${\bf m}_k$ are all tuned and coupled to the NMR coil, so their contributions to $ {\bf B}_\textsc{fb}$ have a common $\psi$, a mild assumption.\cite{gueron_mrm_1991}
Thus, the modified set of Bloch equations in the rotating frame writes as:
    \begin{eqnarray}\label{eq:fb_multilines}
   \nonumber  \dot m_{xi}  &=& -\gamma_{2i}m_{xi}  -\delta\omega_i m_{yi}  +  \omega_1 \sin \psi_1 m_{zi}  - \gamma  m_{zi}  B_\textsc{fb y}\\
   \nonumber  \dot m_{yi}  &=& \delta\omega_i m_{xi}  -\gamma_{2i}m_{yi} - \omega_1 \cos \psi_1 m_{zi}  + \gamma  m_{zi} B_\textsc{fb x}\\
 \nonumber    \dot m_{zi}  &=& - \omega_1 (\sin \psi_1 \, m_{xi}  - \cos \psi_1 \, m_{yi} ) -\gamma_{1i} (m_{zi} - m_{z0i})\\ 
   &+&  \gamma   m_{xi}   B_\textsc{fb y}  -  \gamma  m_{yi}B_\textsc{fb x}
\end{eqnarray}
where  $S_{kx}=\sum_k m_{xk}$, $S_{ky}=\sum_k m_{yk}$, ${\bf B}_\textsc{fb} =G \left [ \cos \psi S_{kx}  +  \sin \psi S_{ky} ,   \cos \psi S_{ky}  -  \sin \psi S_{kx}  , 0\right ]^\dagger$, and $\omega_k$ and $\delta \omega_k$ are the Larmor frequency and offsets of the magnetization components ${\bf m}_k $ in the rotating frame.
Equations \ref{eq:fb_multilines} describe the dynamics of a large dimensional  nonlinear system, and are the basis for the interpretation of our experimental observations.

The radiation damping field  from the probe has an amplitude proportional to the transverse magnetization, to which it lags by 90$^\textsc{o}$.
The induction signal can therefore be used to generate a radiation feedback with controlled phase and gain with respect to the transverse magnetization to control the (nonlinear) dynamics of the spins.
An electronic feedback control unit (eFCU)  was thus built, based on a previous design,\cite{Louis-Joseph_RDsuppress_1995} for the purpose of generating a feedback field controlled in phase and gain.
To this aim, a fraction of the induction signal is picked up through a directional coupler, then demodulated using a 400 MHz local oscillator (LO) reference frequency generated from the $^1$H power amplifier of the spectrometer, filtered and fed back into the probe with adequate phase and gain corrections. Details of the setup are given in the Supplemental Material.\cite{vineeth_thesis}
In particular, this piece of instrumentation is able to invert the effect of radiation damping, and to drive the magnetization towards the south pole of the Bloch sphere.
As explained above, this configuration, together with the longitudinal relaxation process of the spins, allows one to achieve a solution state NMR maser.

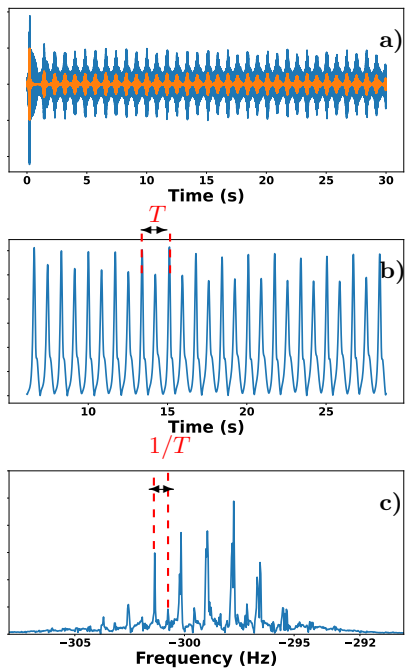
\begin{figure}[h!]
\centering
\subfigure
{\label{fig:methanol_Nopulse_LoONcenter}
\begin{overpic}[width=.3\textwidth]{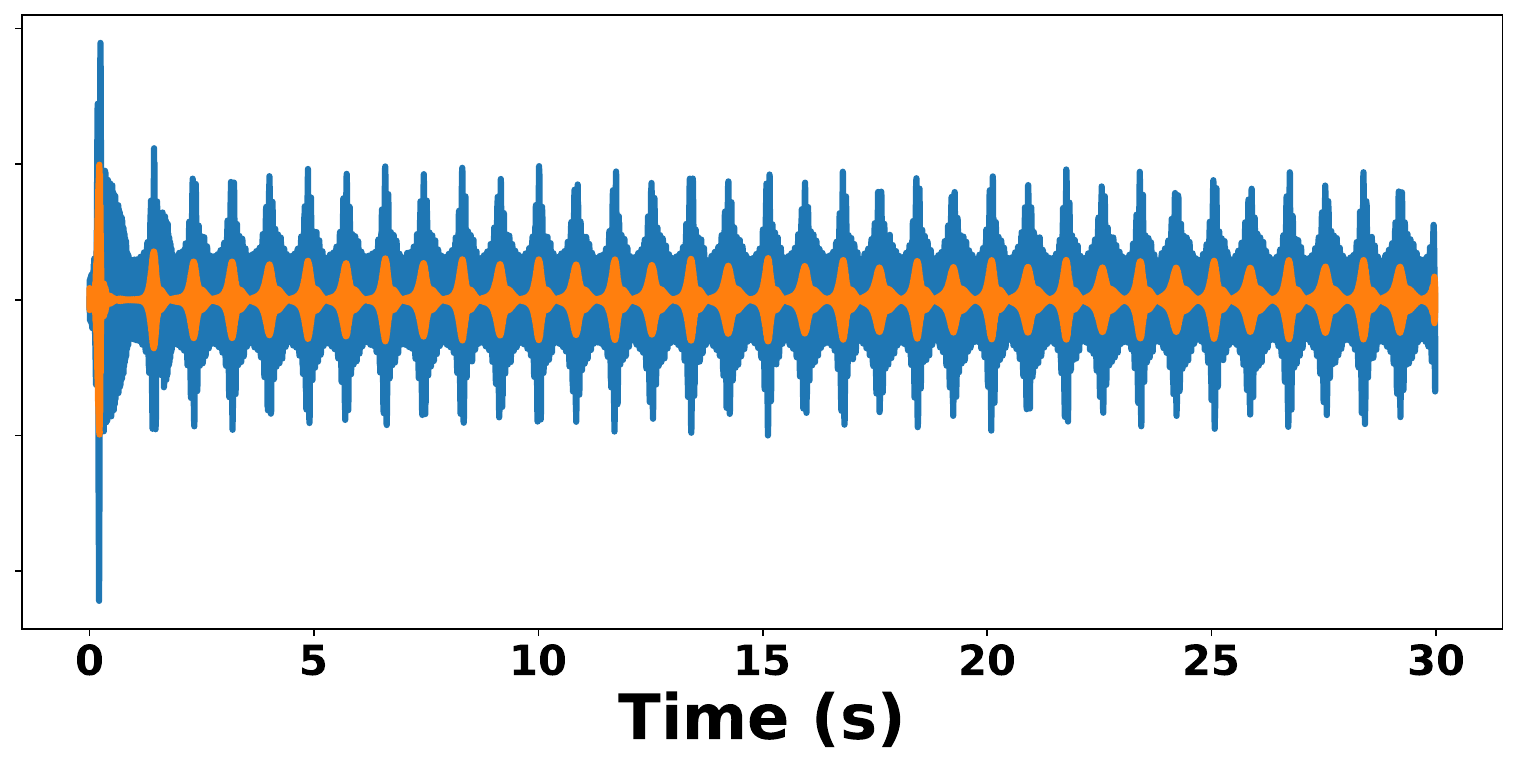}
\put(93,40){{\bf a)}}
\end{overpic}
}

\subfigure { \label{fig:Methanol_limitcycle_signal}
\begin{overpic}[width=.3\textwidth]{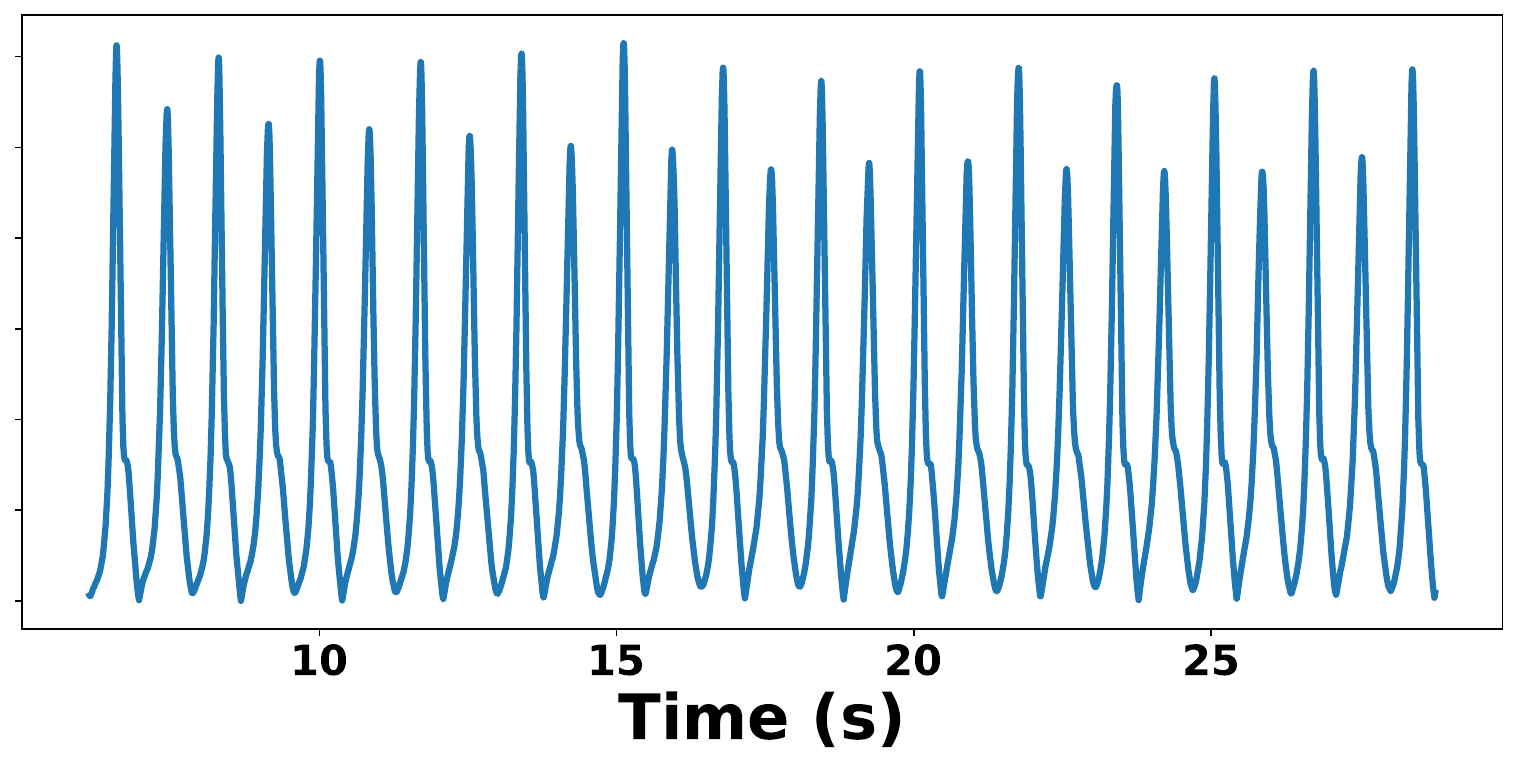}
\put(34,41){\tikz \draw[red,thick,dashed] (0,0)--(0,0.6);}
\put(41,41){\tikz \draw[red,thick,dashed] (0,0)--(0,0.6);}
\put(34,52){\vector(1,0){6.6}}
\put(34,52){\vector(-1,0){0}}
\put(36,54){\color{red}$T$\color{black}}
\put(93,40){{\bf b)}}
\end{overpic}
}

\subfigure {  \label{fig:methanol_driaccomb}
\begin{overpic}[width=.3\textwidth]{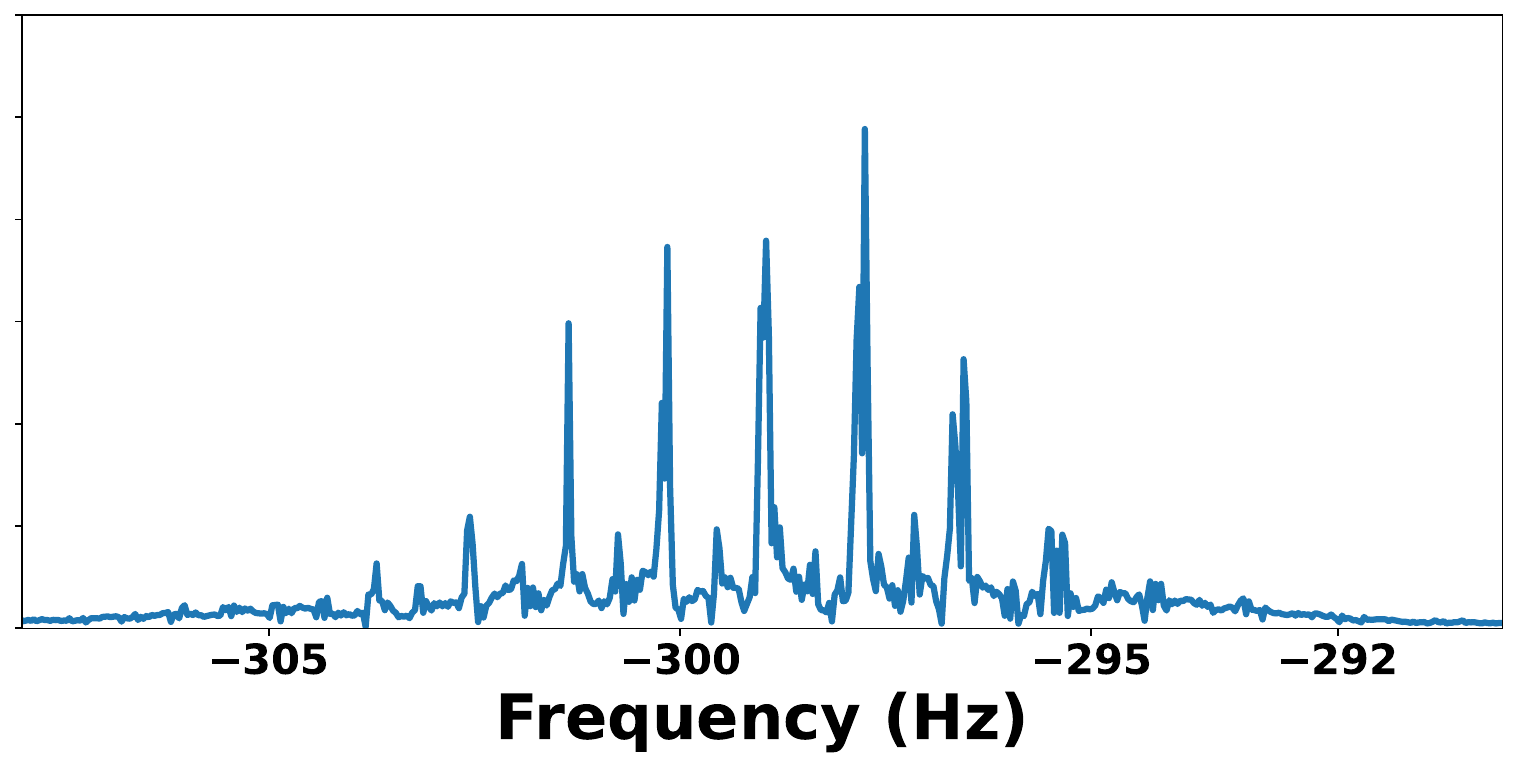}
\put(37,30){\tikz \draw[red,thick,dashed] (0,0)--(0,1.05);}
\put(40.5,15.5){\tikz \draw[red,thick,dashed] (0,0)--(0,1.8);}
\put(38,45){\vector(1,0){4.2}}
\put(37,45){\vector(-1,0){1}}
\put(36,54){\color{red}$1/T$\color{black}}
\put(93,40){{\bf c)}}
\end{overpic}
}
\caption{\label{fig:methanol_single_mode}  Sustained maser obtained with the eFCU-controlled radiation feedback. 
Fig. \ref{fig:methanol_Nopulse_LoONcenter}: sustained \h \, maser signal obtained from a pure methanol sample (blue).   
The induction signal was Fourier transformed, then filtered  around the \h\, CH$_3$ frequency and inverse Fourier transformed to yield the CH$_3$ \h\,maser (orange). 
In \ref{fig:Methanol_limitcycle_signal} the envelope of the CH$_3$ \h\,maser of the last 25 s of the signal in \ref{fig:methanol_Nopulse_LoONcenter} is shown, and the period $T$ indicating the presence of a limit cycle behaviour is  indicated.
The resonance line corresponding to the methanol CH$_3$ \h\,maser signal has a complex pattern resulting from the periodic nature of the NMR signal. The particular shape of the envelope yields a ensemble of spectral lines separated by $1/T$.
Experiments were performed on a 400 MHz Bruker spectrometer equipped with triple resonance 5 mm (TXI) probe.}
\end{figure}

As a first implementation of our radiation feedback  control device, experiments were performed on a sample containing a 1:1 mixture of methanol and water (225 $\mu$L each, 50 $\mu$L of deuterated DMSO) 
Moreover,  20 $\mu$L of a 200 mM CUSO$_4$ solution were added to increase transverse relaxation.
As a consequence, the CH$_3$ and OH resonance multiplets were unresolved on the one-dimensional spectrum (see  Supplemental Material for a reference NMR spectrum).
The local oscillator (LO) frequency that serves for the quadrature modulation and demodulation in the eFCU was set roughly halfway between the CH$_3$ and OH  \h \, resonance frequencies (-298.3 Hz and 295.4 Hz from CH$_3$ and OH, respectively), and the gain and phase of the feedback device were set to rotate the magnetization towards the $-z$ direction.
The signal obtained in these conditions exhibited initially a few maser bursts of decaying intensities followed by a series of regularly spaced, non attenuated, maser pulses (Fig.  \ref{fig:methanol_Nopulse_LoONcenter}).
Induction signals could typically be observed for tens of seconds, and actually maintained "indefinitely" without any damping of the pulse envelope.
In this experiment, no pulse was applied to initiate the maser sequence, so that the train of maser pulses occurs after a delay of several tens of ms.
This suggests that the sequence is triggered by the circuit noise,\cite{augustine_2000} since the equilibrium direction is unstable for this set up, or less likely by the weak rf field ($4$ Hz in this case) leaking from the eFCU device (see SM).

The Fourier transform of the signal contains a component at the CH$_3$ frequency, but also an additional resonance line corresponding to its image generated by the demodulation/remodulation stages of the eFCU. Details are given in the SM. 
The induction signal corresponding to the CH$_3$ is therefore extracted from the spectrum by  selecting the resonance line of interest and filtering out the image peak and the rest of the spectrum.
Applying an inverse Fourier transform then  leads to the pure CH$_3$ maser induction signal, shown as the orange trace in Fig.\ref{fig:methanol_Nopulse_LoONcenter}.

Note that in these experiments, no maser was associated with the OH resonance, which was ascribed to the presence of chemical exchange with water protons.\cite{mao_1997,rodriguez_2002}.
In this case, spin jumping from a site to another induces a loss of transverse coherence in a time shorter than the characteristic radiation feedback time, thereby limiting  its efficiency.
The absence of an OH signal  in this maser experiment was confirmed by repeating the same experiment with slightly shifted LO frequencies.
Changes of the LO frequency shifted the CH$_3$ signal image accordingly whilst no additional line at the OH frequency, nor image thereof, were observed (see Fig. \ref{fig_SI:Methanol_position_LO} in the SM).
The CH$_3$ spectrum, shown in Fig.\ref{fig:methanol_driaccomb}, is composed of  series of lines with alternating intensities and separated by the frequency $1/T$, as  expected from the Fourier transform of a periodic signal of period $T$.
It is simply interpreted by remarking that in the case at hand, the NMR signal can be decomposed into two nearly identical waveforms of the same period and shifted in time. This aspect is expanded in Eqs.\ref{eq_SI:periodicsignal_1}-\ref{eq_SI:spec_periodic_2} of the Supplemental Material.

The steady-state periodic evolution of the detected signal is a striking feature of this experiment.
Indeed, instead of the expected decay of the amplitude of the transverse magnetization towards a stationary value predicted by the Maxwell-Bloch equations,  and observed experimentally,\cite{Bosiger1977, suefke_2017, Weber2019, appelt_jmr_2021, hope2021} the envelope of the transverse magnetization does not reach a constant value but  oscillates indefinitely, 
This asymptotic behaviour is the signature of a limit cycle of the dynamical system\cite{holmes} and implies the existence of a non hyperbolic fixed point of the three-dimensional Maxwell-Bloch equations (i.e., with pure imaginary eigenvalue).
However, this is possible only if  $\gamma_1=0$, in which case the exact analytical solution shows that the magnetization simply evolves towards the z axis,\cite{bloom57} without the possibility of multiple masers. 
Further details are given in the Supplemental Material.
These experiments therefore indicate a breakdown of the Maxwell-Bloch equations, and their interpretation requires a more complex description.

\begin{figure}
\centering
\subfigure{\label{fig:multimode_inhomogenity_fid}
\begin{overpic}[width=.45\textwidth]{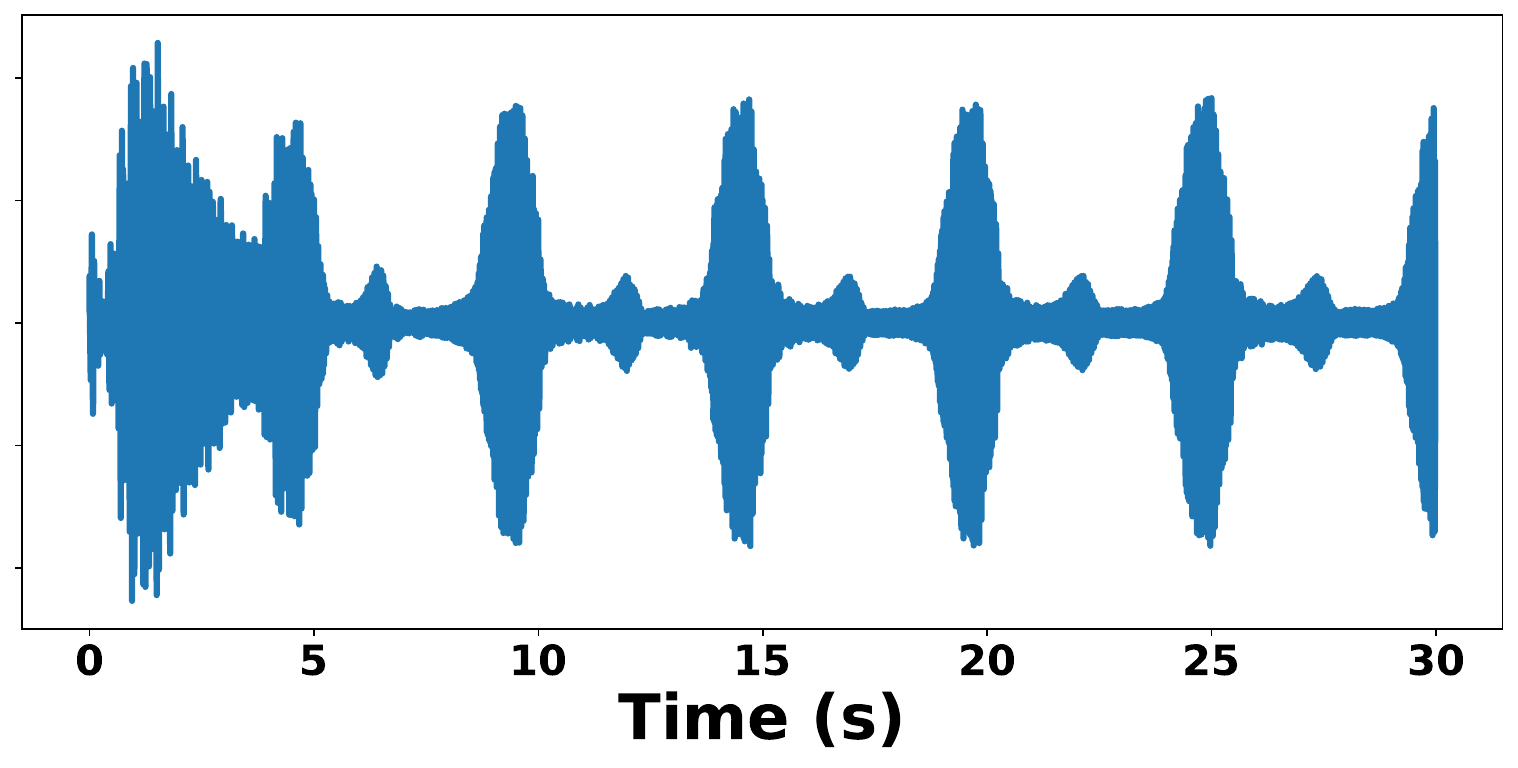}
\put(95,14){a)}
\end{overpic}
}
\subfigure{\label{fig:multimode_inhomogenity_spec}
\begin{overpic}[width=.45\textwidth]{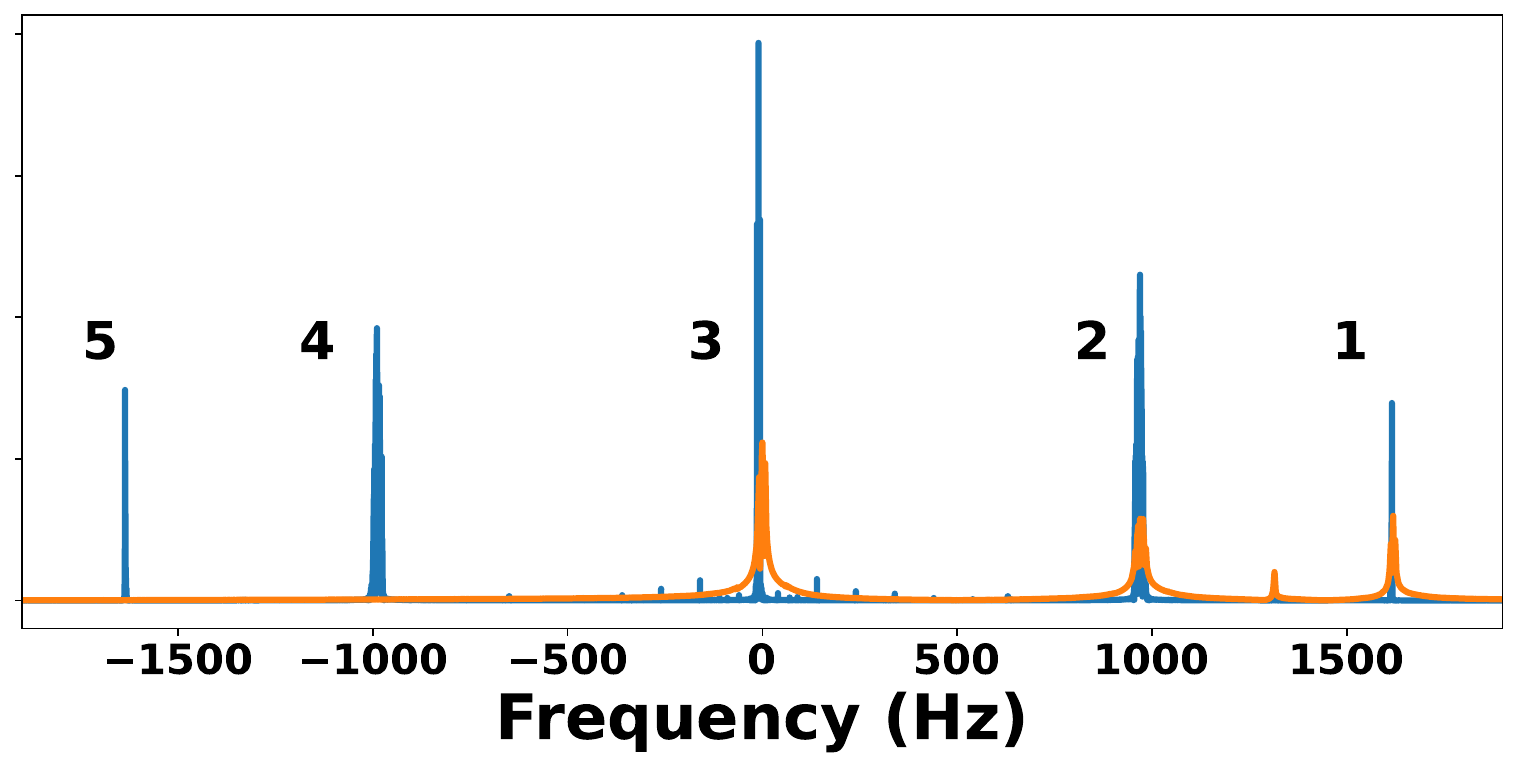}
    \put(87,10){\tikz \draw[orange,thick,dashed,rounded corners] (0,0) rectangle (0.7,3);}
    \put(88,40){{\bf OH}}
    \put(71,10){\tikz \draw[orange,thick,dashed,rounded corners] (0,0) rectangle (0.7,3);}
    \put(71,40){{\bf CH$_2$}}
    \put(45,10){\tikz \draw[orange,thick,dashed,rounded corners] (0,0) rectangle (0.7,3);}
    \put(45,40){{\bf CH$_3$}}
\put(95,14){b)}
\end{overpic}
} 

\subfigure{\label{fig:multimode_inhomogenity_ch3}
\begin{overpic}[width=.3\textwidth]{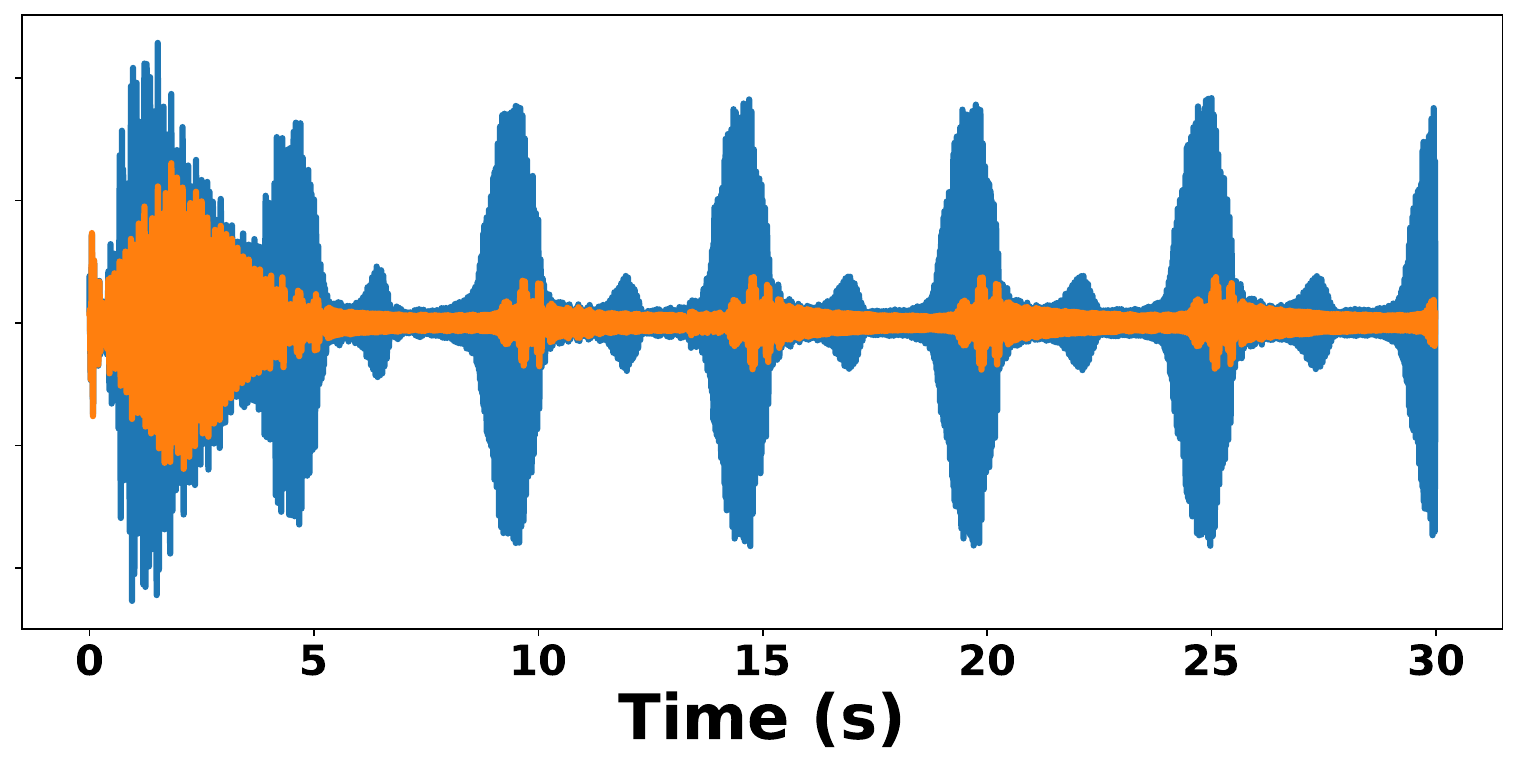}
\put(81,43){{\bf CH$_3$}}
\put(90,12){c)}
\end{overpic}
}
\subfigure{\label{fig:multimode_inhomogenity_ch2}
\begin{overpic}[width=.3\textwidth]{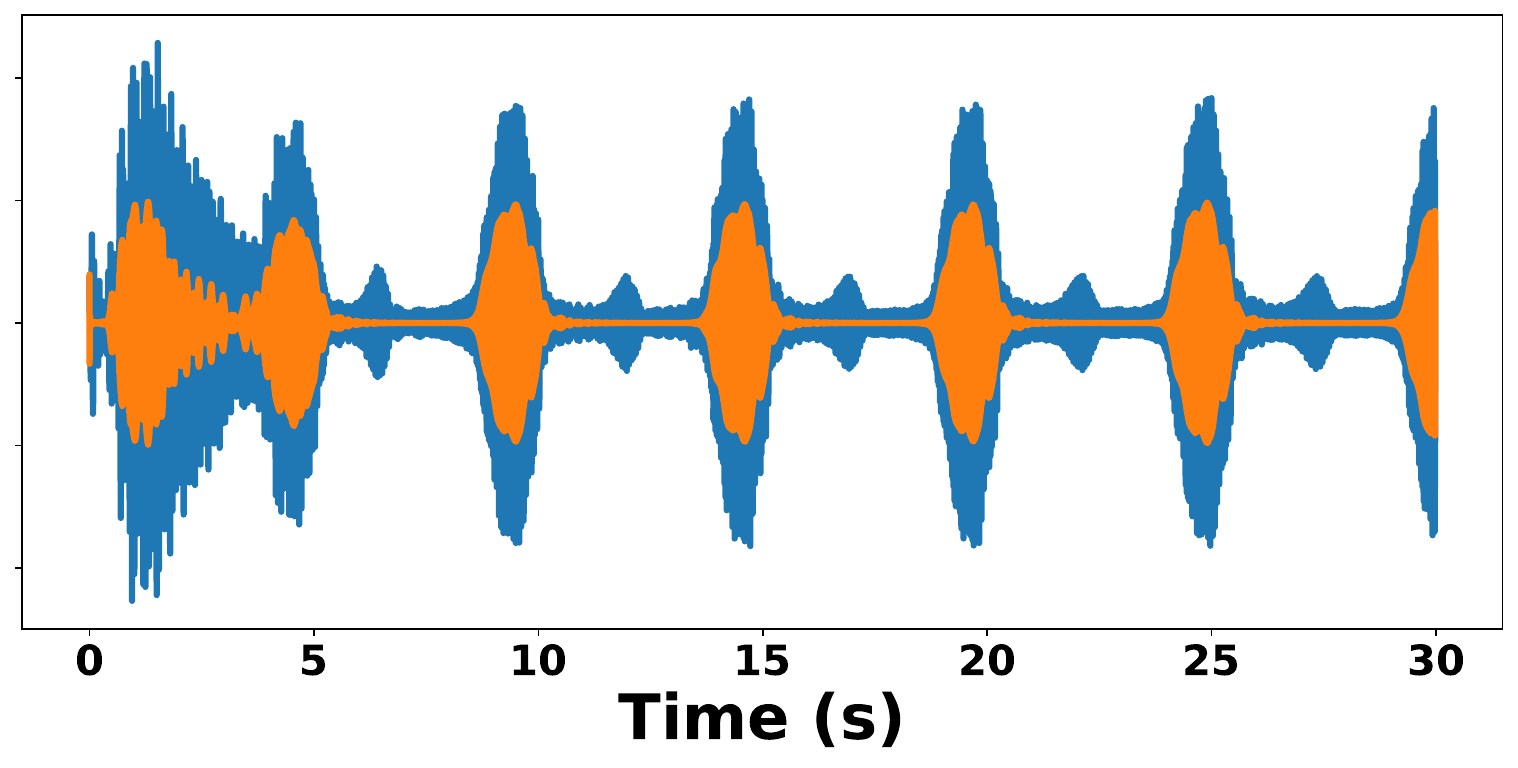}
\put(81,43){{\bf CH$_2$}}
\put(90,12){d)}
\end{overpic}
}
\subfigure{\label{fig:multimode_inhomogenity_oh}
\begin{overpic}[width=.3\textwidth]{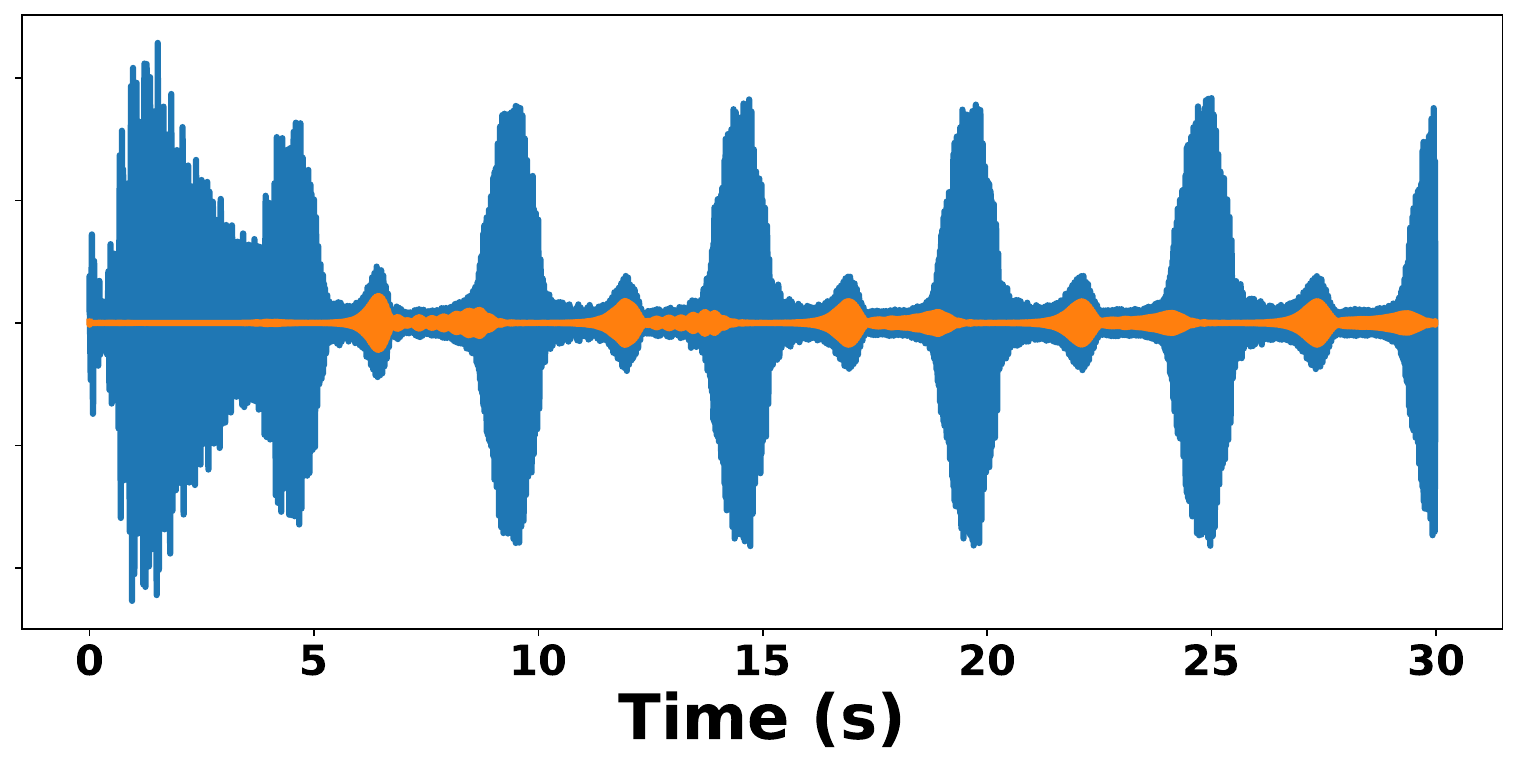}
\put(81,43){{\bf OH}}
\put(90,12){e)}
\end{overpic}
}
\caption{\label{fig:multimode_inhomogenity}  Sustained multimode maser of ethanol generated by the eFCU. A multiple maser signal acquired for 30 s is shown in  \ref{fig:multimode_inhomogenity_fid}.  The associated spectrum is shown in \ref{fig:multimode_inhomogenity_spec}, where peaks 1-3 correspond to OH, CH$_2$, CH$_3$  (dashed lined orange boxes). 
 Mirror peaks of 1 and 2 are labelled 5 and 4, respectively, and are due to quadrature demodulation and re-modulation (see text for details).  The 1D spectrum of ethanol obtained in the absence of eFCU is depicted in orange (see also Fig. \ref{fig_SI:Methanol_ethanol_Ref} in SM).
In \ref{fig:multimode_inhomogenity_ch3}-\ref{fig:multimode_inhomogenity_oh}, the inverse Fourier transform of the regions corresponding to the CH$_3$, CH$_2$ ans OH resonance lines of the full spectrum are shown in orange, superimposed with the total induction signal. 
} 
\end{figure}
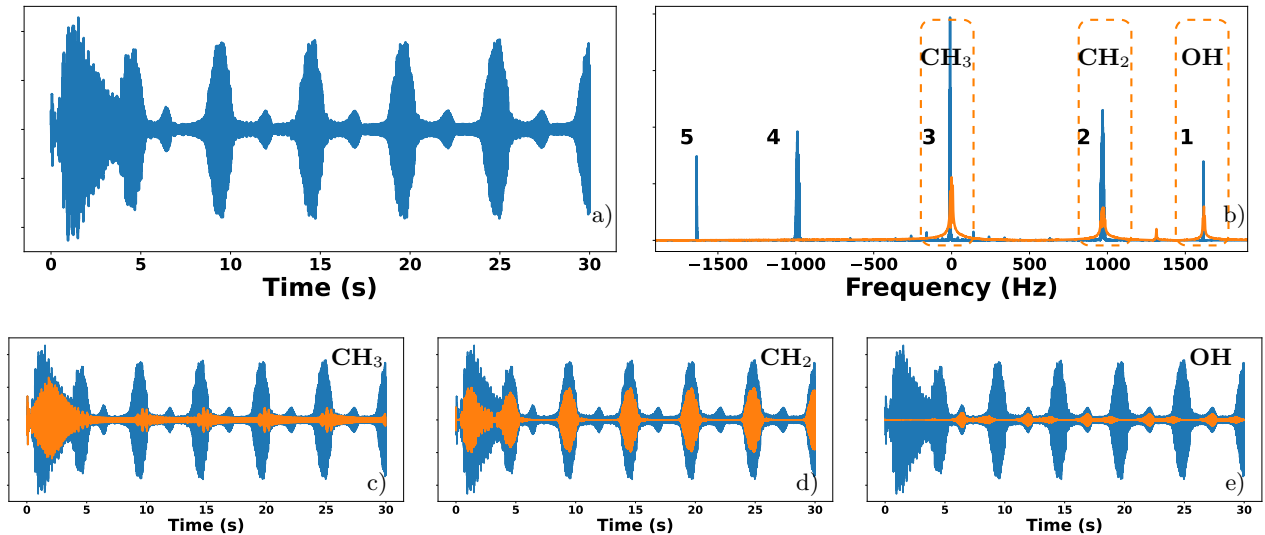

Sustained maser experiments were also performed on a sample of pure ethanol (450 $\mu$L of ethanol + 50 $\mu$L of deuterated DMSO).
Again, the phase and gain of the radiation feedback control device were set so as to invert radiation feedback and to drive the magnetization towards the $-z$ direction.
The local oscillator frequency  was set to the center of the CH$_3$ multiplet in these experiments.
The resulting train of sustained maser pulses is shown in Fig. \ref{fig:multimode_inhomogenity}.
This signal was analyzed through its Fourier transform, shown in Figure \ref{fig:multimode_inhomogenity_spec}.
The spectrum exhibits resonance lines corresponding respectively  to the OH, CH$_2$, and CH$_3$ \h\, resonances of ethanol (labelled 1, 2, 3).
Peaks labelled 4 and 5 are the  respective mirror images  of the CH$_2$ and OH signals (labelled 1 and 2 in the figure) which, as explained above, are expected, and are due to the demodulation and re-modulation  stages with the reference LO frequency.
The latter being set on the CH$_3$ frequency, no such mirror peak exists for this resonance line.   
Again, the raw induction signal shows a superposition of regularly spaced and non attenuated maser bursts.
This limit cycle kind of behaviour is yet a further illustration the breakdown of the Maxwell-Bloch equations.
Disentangling the specific contributions of the different spins from the total signal was achieved in a straightforward manner by back-calculating the time domain signals relative to each resonance line through selective inverse Fourier transform restricted to the spectral region of interest.
Each of the reconstructed selective induction signals exhibits a series of sustained maser pulses, showing that this experiment produces a three-mode maser built on the radiation feedback of three resolved lines.
Moreover, it is found that each signal evolves toward a periodic steady state maser, meaning that each spin moiety obviously undergoes a stationary limit cycle, as in the case of methanol.
These are depicted in Figure \ref{fig:multimode_inhomogenity}.

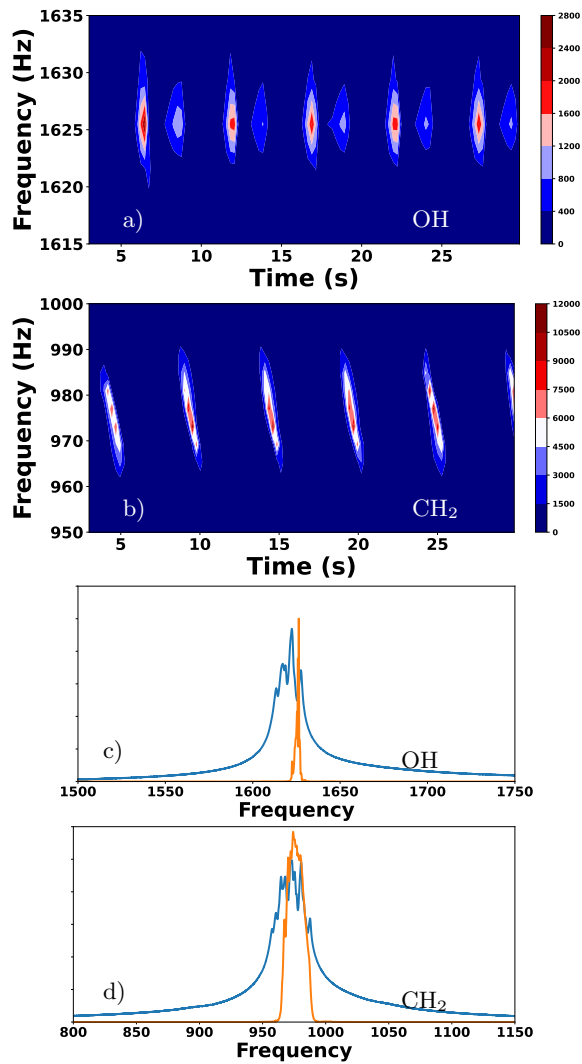
\begin{figure}
    \centering
\subfigure{\label{fig:timefreq_oh}\begin{overpic}[scale=0.3]{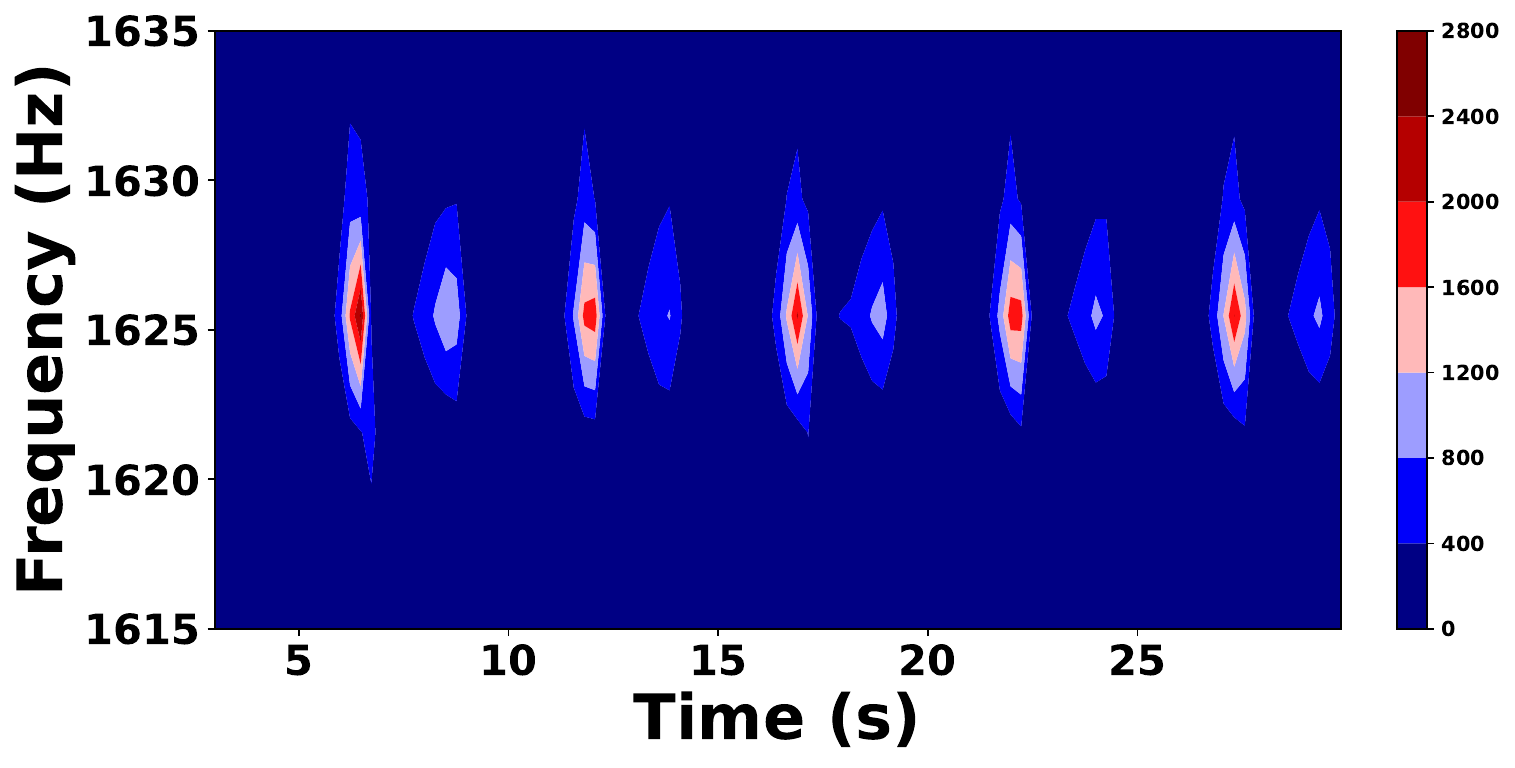}
    \put(70,12){\color{white} OH \color{black}}
    \put(20,12){\color{white} a) \color{black}}
\end{overpic}}\\
\vspace{-0.4cm}
\subfigure{\label{fig:timefreq_ch2}\begin{overpic}[scale=0.3]{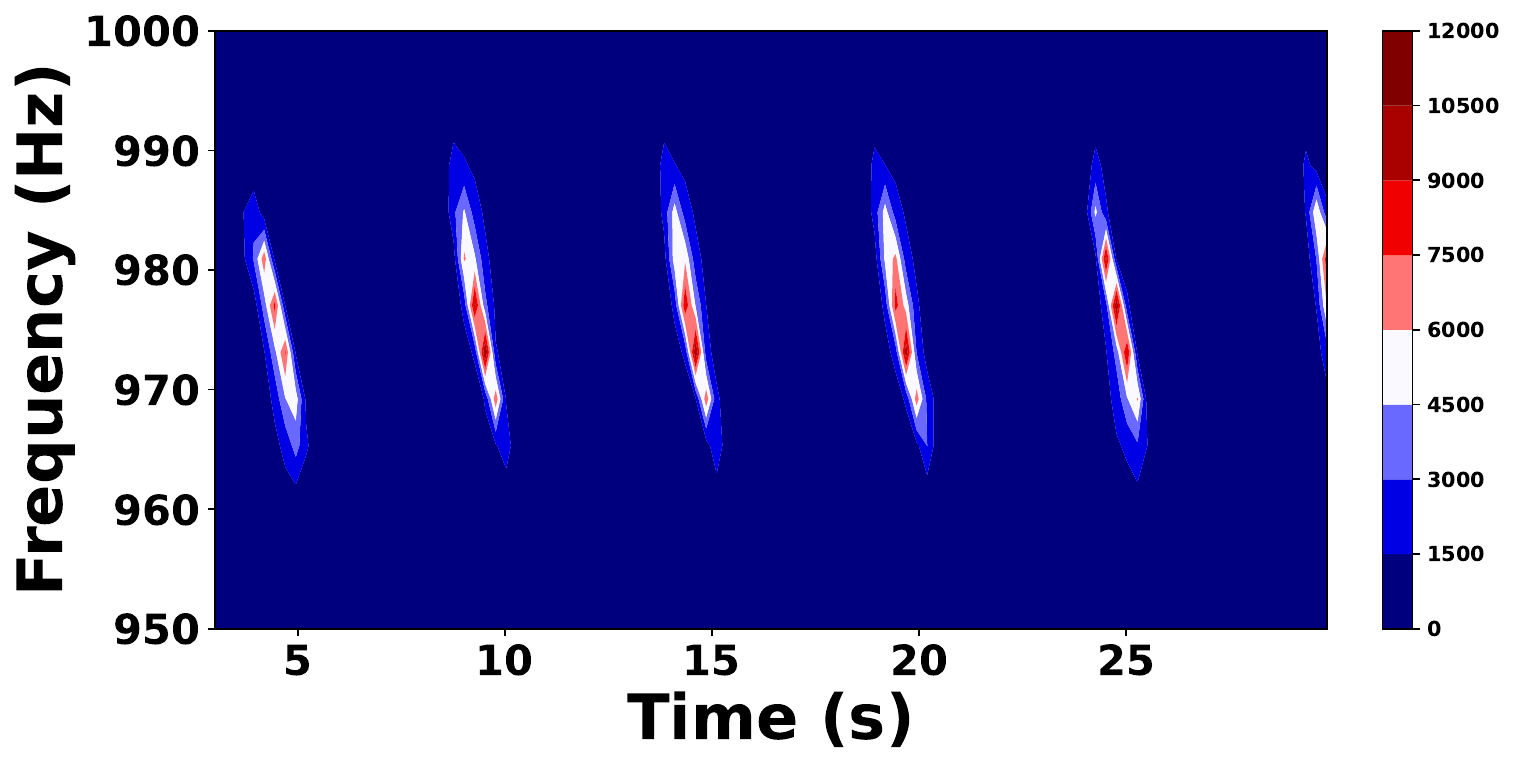}
    \put(70,12){\color{white} CH$_2$ \color{black}}
    \put(20,12){\color{white} b) \color{black}}
\end{overpic}}\\   
\vspace{-0.4cm}
\subfigure{\label{fig:clustfreq_oh}\begin{overpic}[scale=0.25]{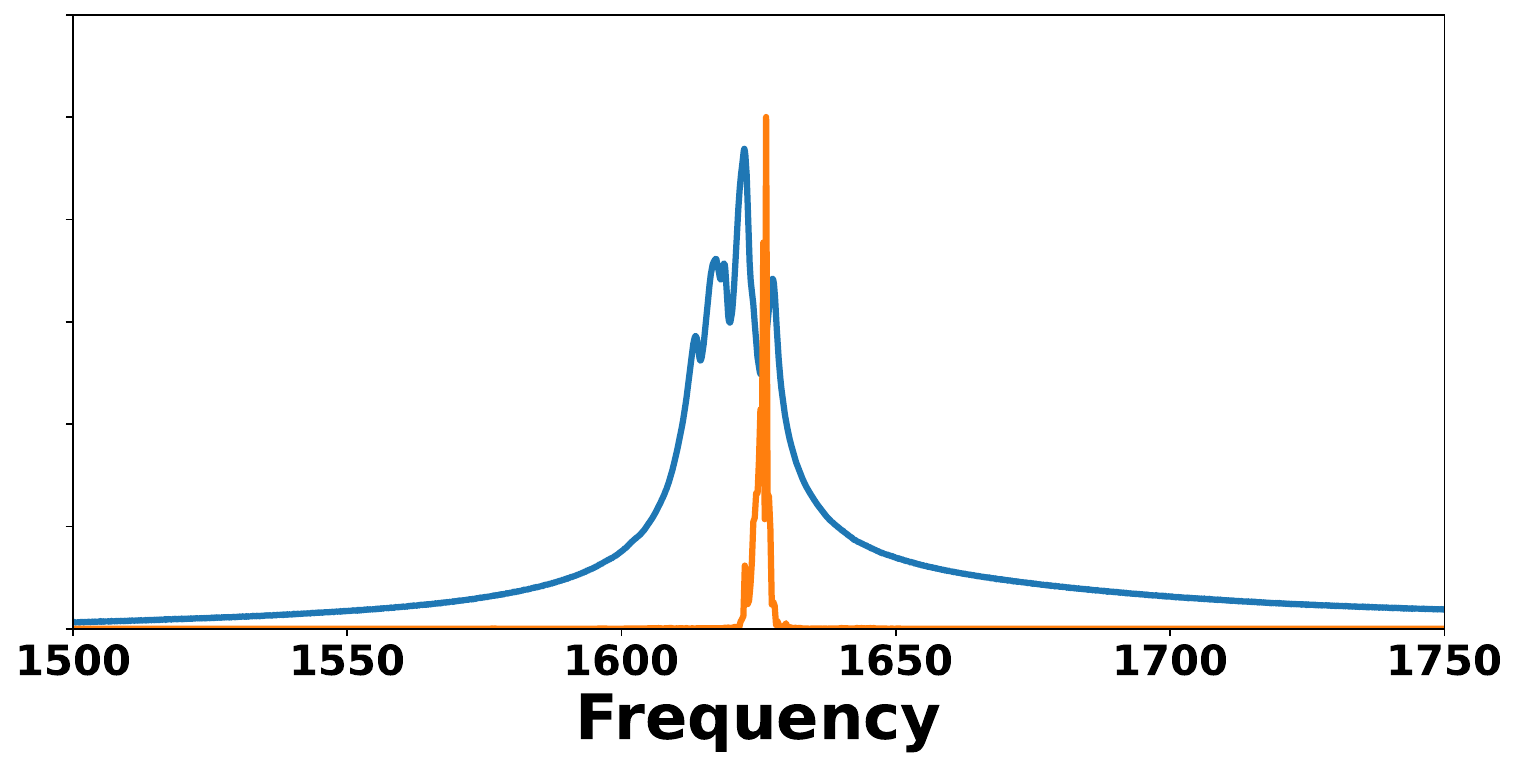}
    \put(70,12){ OH}
    \put(10,14){c) }
\end{overpic}}  \\
\vspace{-0.4cm}
\subfigure{\label{fig:clustfreq_ch2}\begin{overpic}[scale=0.25]{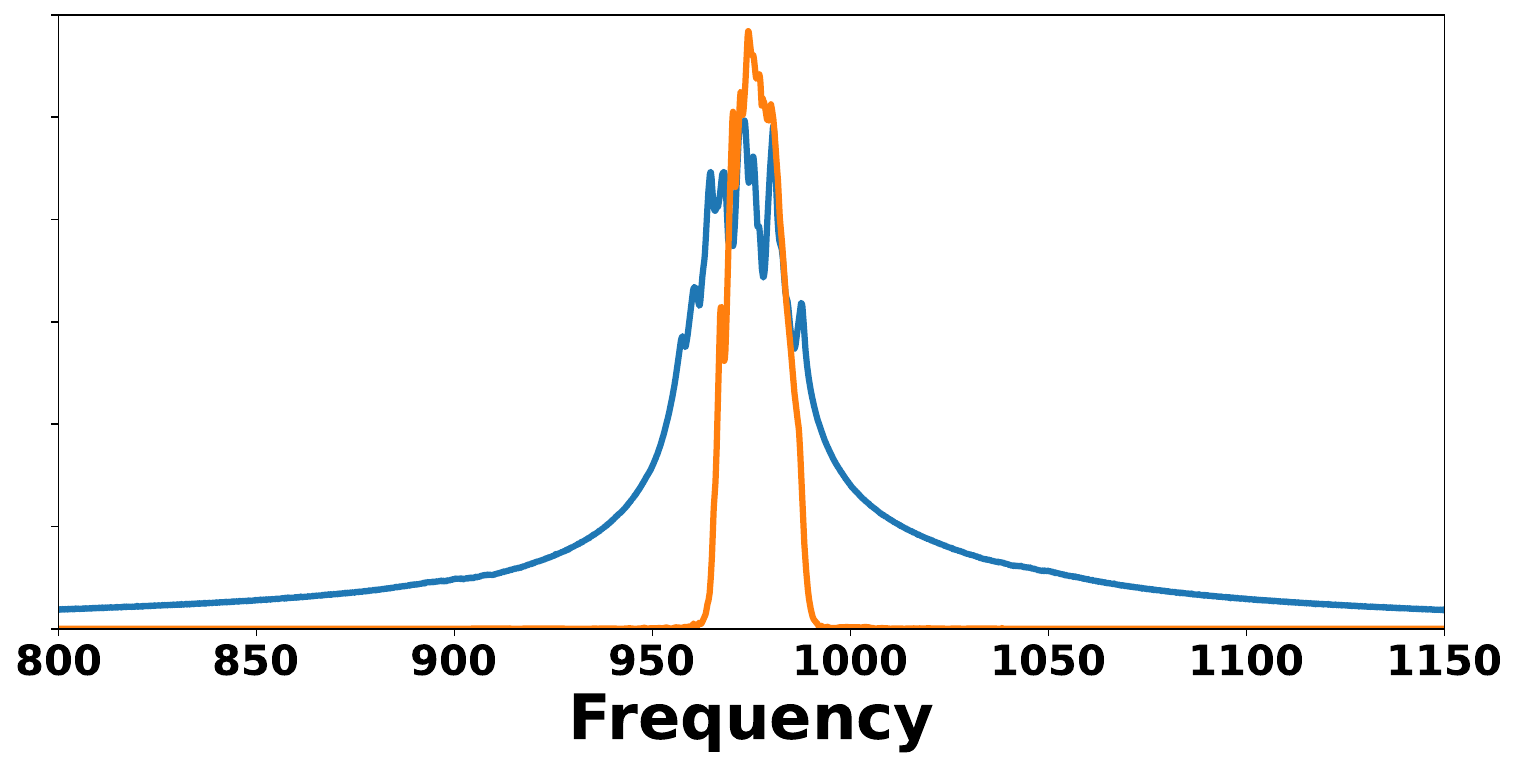}
    \put(70,12){ CH$_2$}
    \put(10,14){d) \color{black}}
\end{overpic}}   

    \caption{\label{fig:timefreq_ethanol} Spectral clustering in the three-mode ethanol maser. The time envelopes shows a succession of pairs of maser pulses in a narrow frequency band for OH  (\ref{fig:timefreq_oh}), and several maser components shifted in time for CH$_2$ (\ref{fig:timefreq_ch2}). The Fourier transform of a single maser pulse (\ref{fig:clustfreq_oh}-\ref{fig:clustfreq_ch2}) exhibits narrower  spectral clustering motif for OH (14.46 Hz) than for CH$_2$ (29.81 Hz). The magnitude spectra are shown in orange, and superimposed to the 1D spectra (in blue).}
\end{figure}

A time-frequency analysis of the maser induction decays reveals the composite structure of the maser pulses that results from the different magnetization components of the multiplets.
In the case of the CH$_2$  mode of the maser, the maser pulses contain several resonance frequencies, the contributions of which to the maser are shifted in time.
In contrast, the OH maser mode a single frequency is identified, and the shape of the signal is reminiscent of the methanol case, with  maser bursts alternatively larger and smaller.
Results are depicted in Figs. \ref{fig:timefreq_oh}-\ref{fig:timefreq_ch2}.
Besides, the Fourier transform of a single maser burst shows that frequencies tend to cluster in a narrow region of the spectrum that does not reflect the actual multiplet structure and width observed on a simple 1D spectrum.
This spectral clustering is a consequence of the nonlocality of the feedback field in the frequency domain expressed by Eq. \ref{eq:fb_multilines}, and can be qualitatively understood by realizing that when the total feedback field dominates the frequency spread of the multiplet, the difference of offset becomes negligible, and the moments tend to undergo the same effective field. 
These effects are qualitatively well reproduced by simulations (see Fig. \ref{fig_SI:Ethanol_lineProfile_OH} and below).

In order to understand the inadequacy of the Maxwell-Bloch model for a single magnetization with our expriments, we investigated the simplest version of Eqs. \ref{eq:fb_multilines} consisting of two moments ${\bf m}_{1,2}$ subject to radiation feedback.
Simulations of this toy model were performed for identical moments $m_0$ such that $G m_{0}=7.0$ Hz, relaxation rates $\gamma_1=0.2 \text{ s}^{-1}$ and $\gamma_2=5 \text{ s}^{-1}$, and an initial magnetization flip angle = $10^{o}$.
The phase of the feedback field was set to $\psi=+\pi/2$.

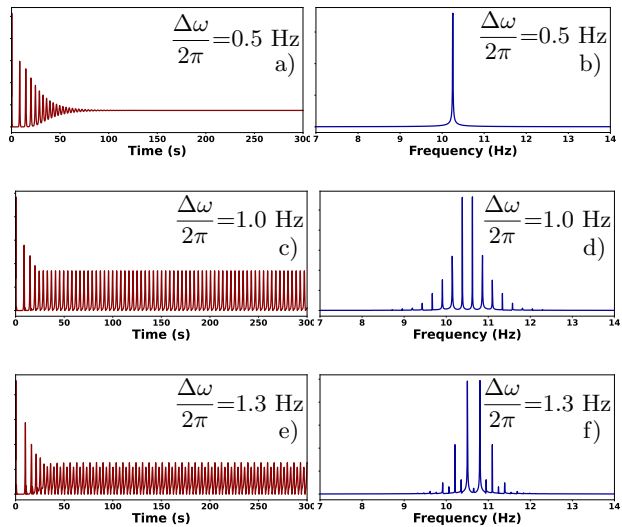
\begin{figure}[h!]
\centering
\subfigure{\label{fig:linesep_sig_201}
\begin{overpic}[scale=.16]{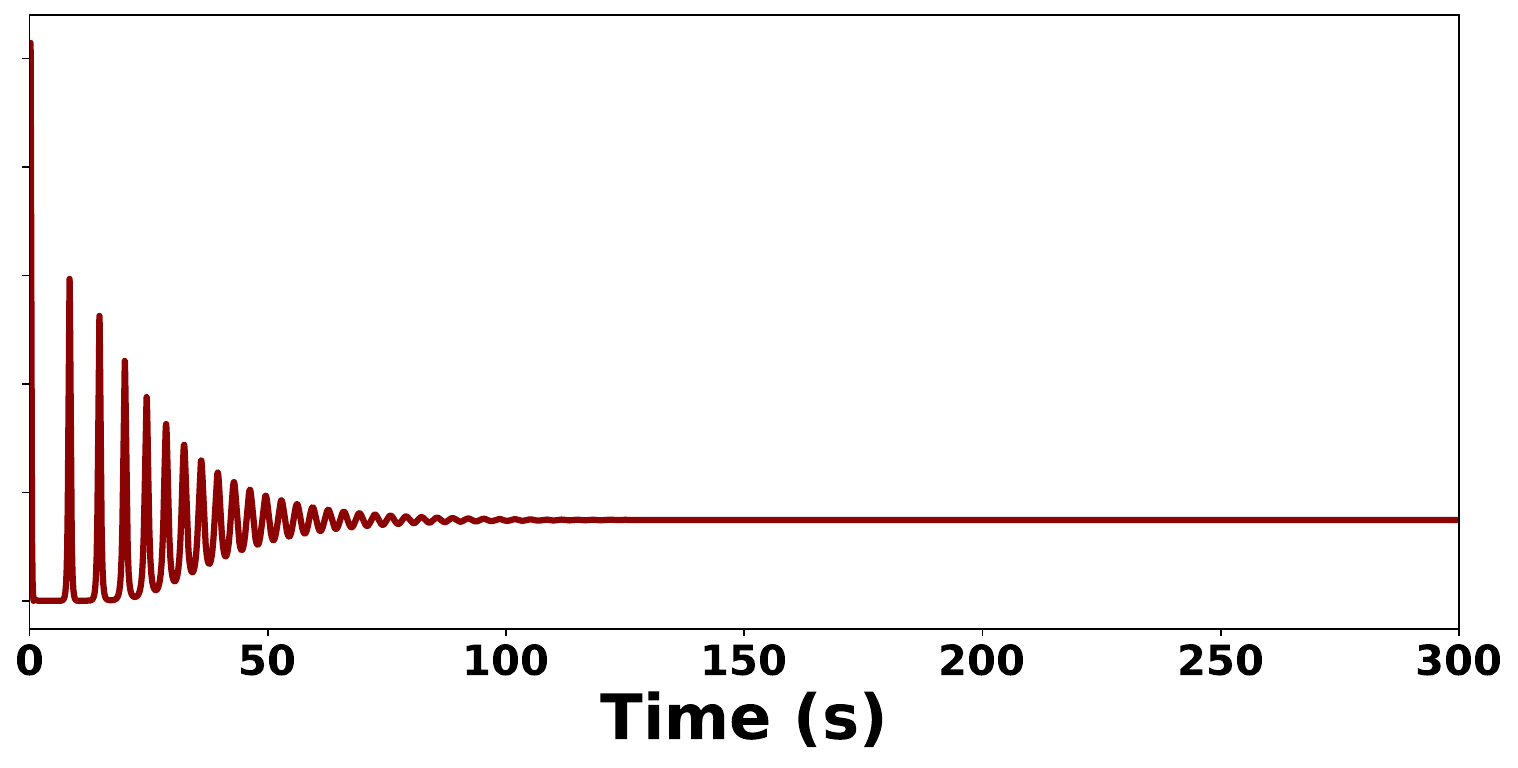}
    \put(53,37){$\displaystyle \frac{\Delta\omega}{2\pi}$=0.5  Hz}
    \put(87,29){a)}
\end{overpic}
}
\hspace{-0.5cm}
\subfigure{\label{fig:linesep_spec_201}
\begin{overpic}[scale=.16]{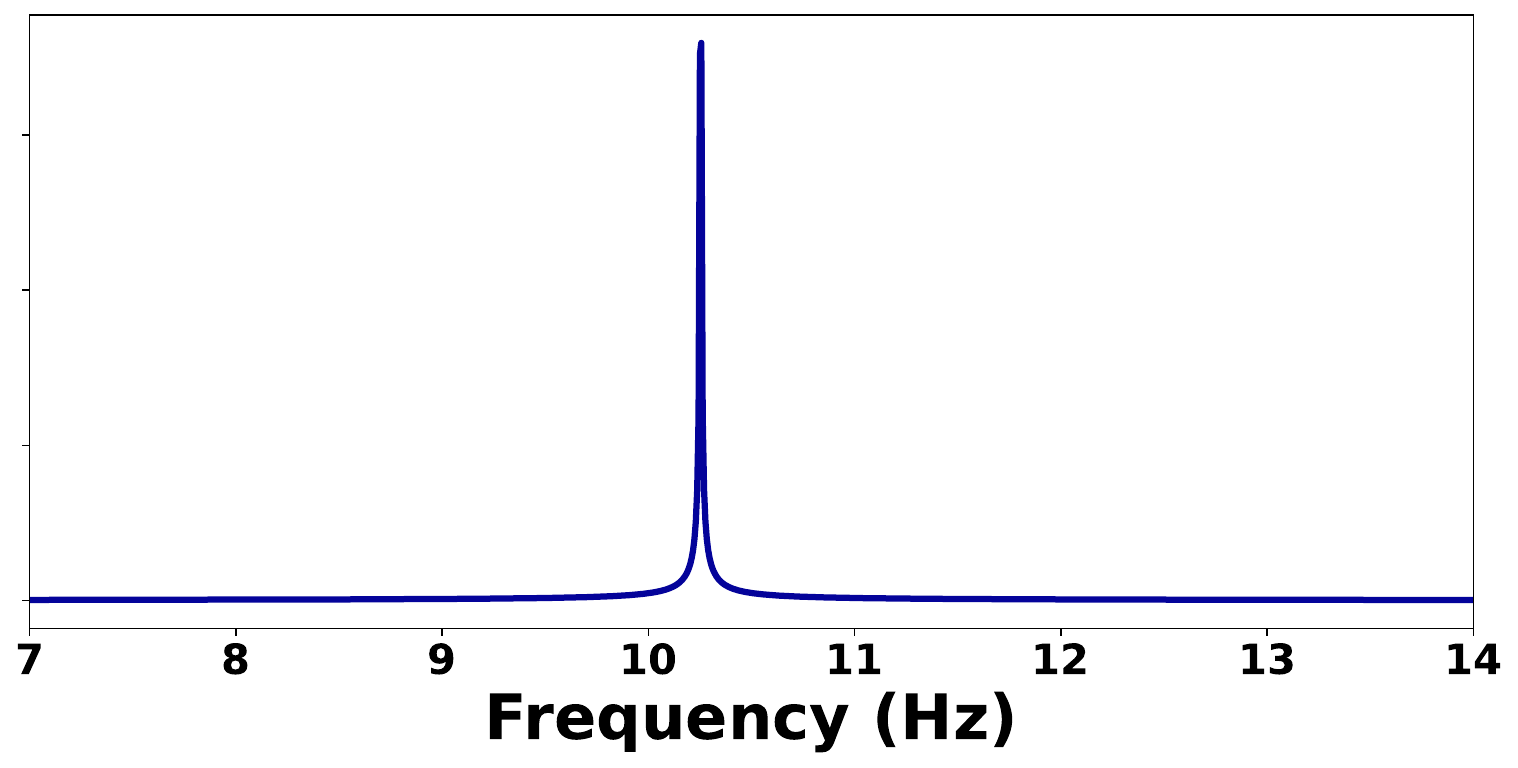}
    \put(54,38){$\displaystyle \frac{\Delta\omega}{2\pi}$=0.5 Hz}
    \put(87,29){b)}
\end{overpic}
}

\subfigure{\label{fig:linesep_sig_203}
\begin{overpic}[scale=.16]{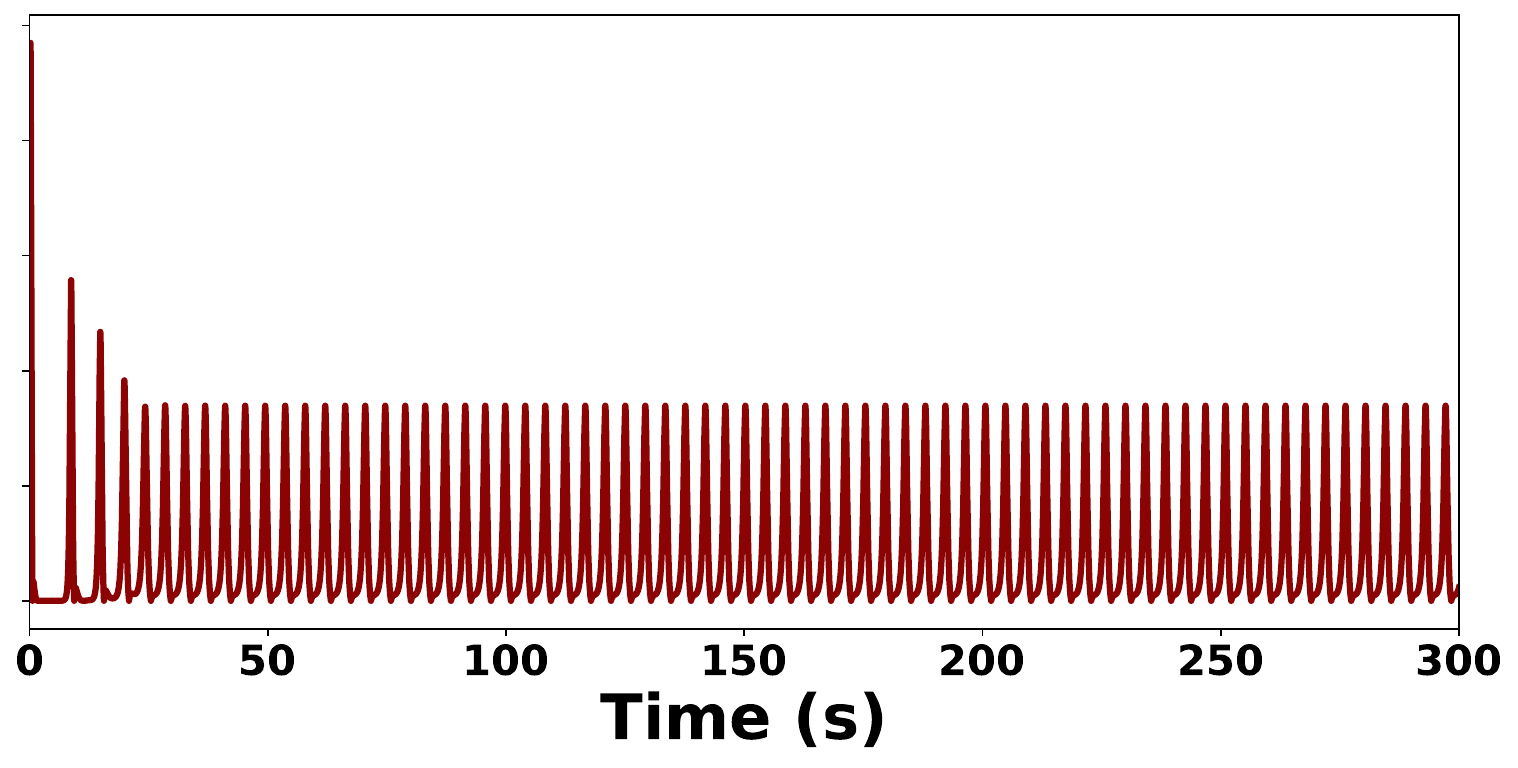}
    \put(53,38){$\displaystyle \frac{\Delta\omega}{2\pi}$=1.0 Hz}
    \put(87,29){c)}
\end{overpic}
}
\hspace{-0.5cm}
\subfigure{\label{fig:linesep_spec_203}
\begin{overpic}[scale=.16]{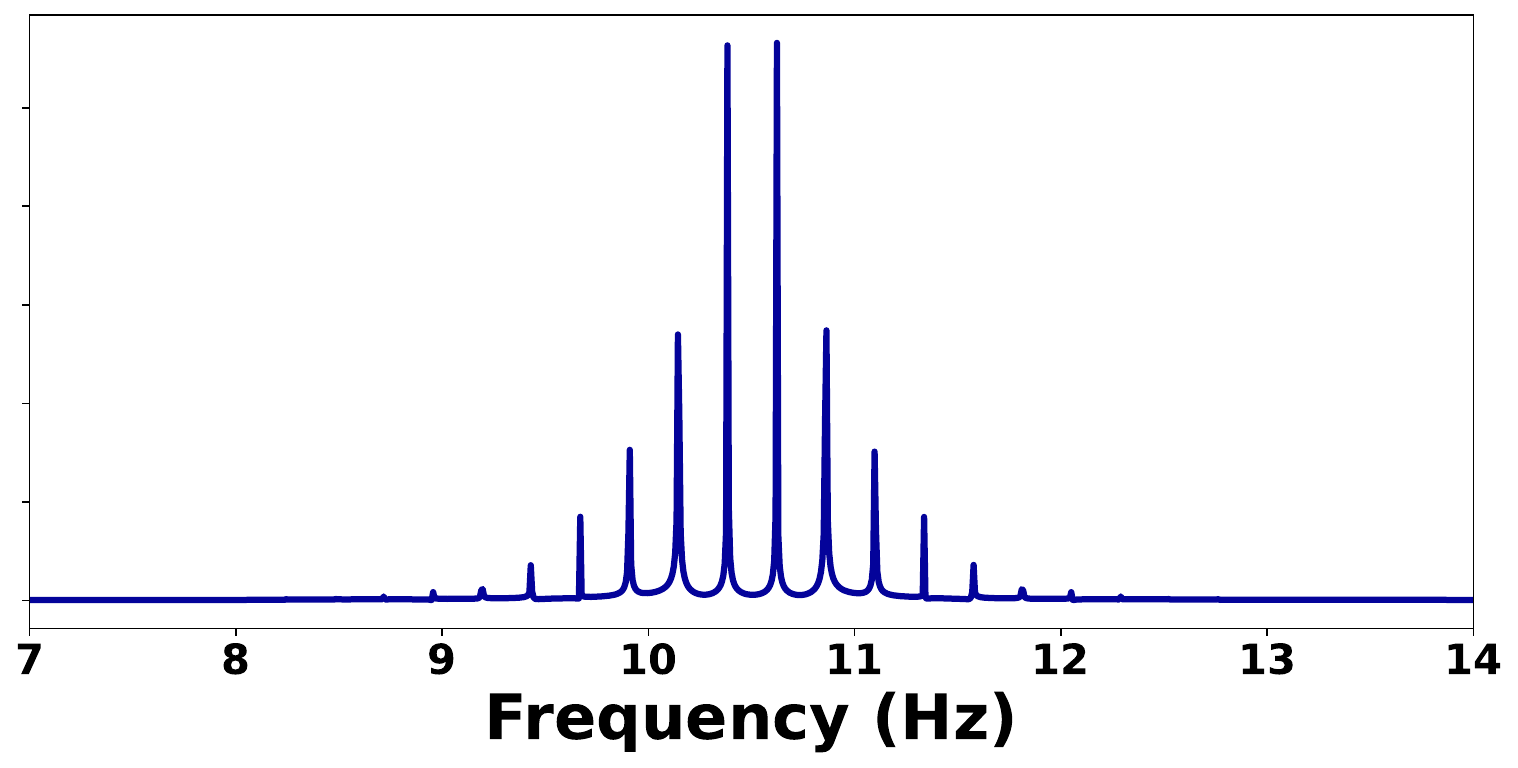}
    \put(54,38){$\displaystyle \frac{\Delta\omega}{2\pi}$=1.0 Hz}
    \put(87,29){d)}
\end{overpic}%
}

\subfigure{\label{fig:freqDoub_signal}
\begin{overpic}[scale=.16]{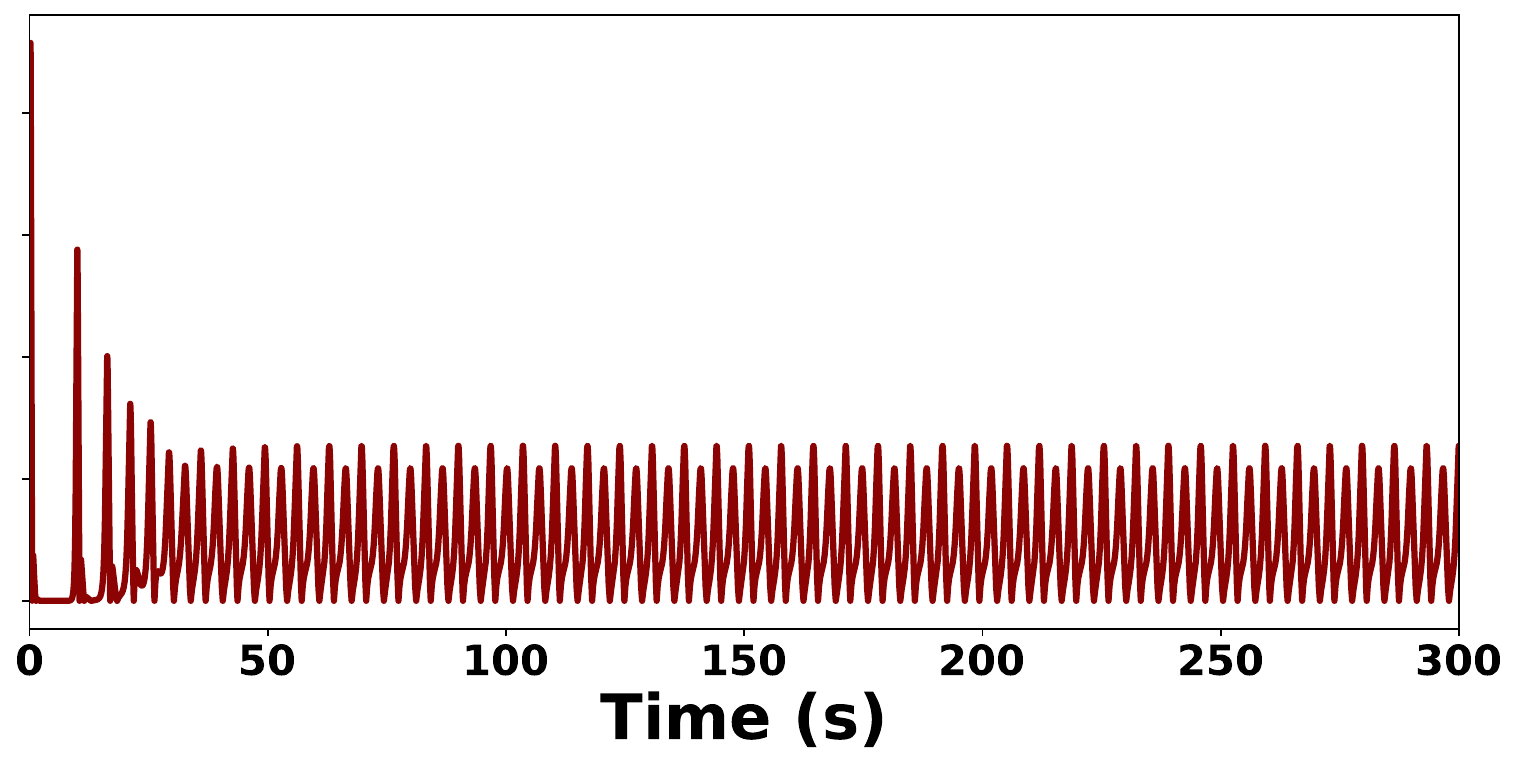}
    \put(53,38){$\displaystyle \frac{\Delta\omega}{2\pi}$=1.3  Hz}
    \put(87,29){e)}
\end{overpic}
}
\hspace{-0.5cm}
\subfigure{\label{fig:freqDoub_spectrum}
\begin{overpic}[scale=.16]{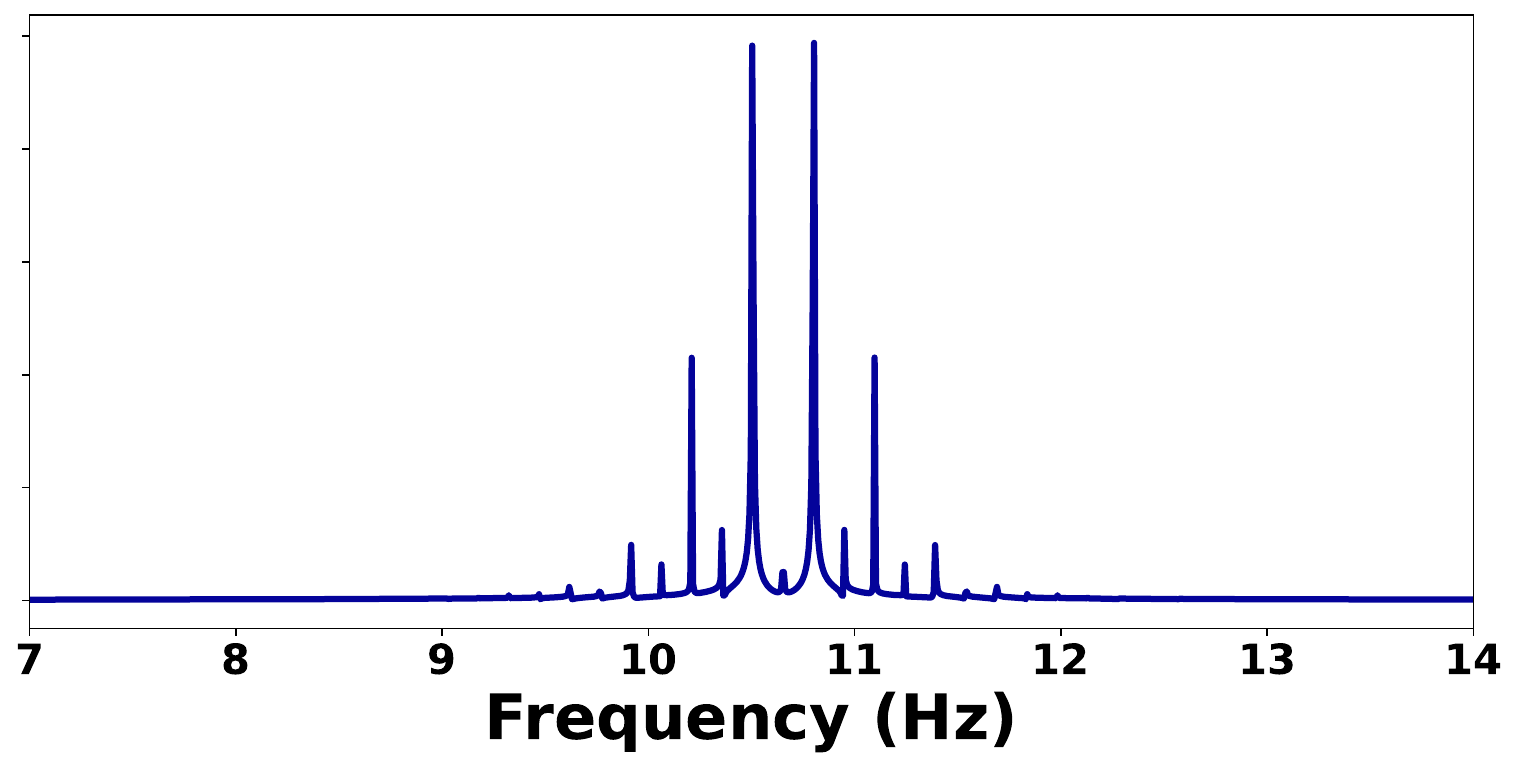}
    \put(54,38){$\displaystyle \frac{\Delta\omega}{2\pi}$=1.3 Hz}
    \put(87,29){f)}
\end{overpic}%
}
\caption{\label{fig:lineseparation_noB1_notresolved} 
Amplitude of $m(t)$ and Fourier spectrum of sustained masers for a nonlinear feedback model with two moments. 
In  \ref{fig:linesep_sig_201} ($\Delta\omega/2\pi = 0.5$ Hz) the magnetization dynamics is analogous to the one predicted by the Maxwell-Bloch equations.
In  \ref{fig:linesep_sig_203} ($\Delta\omega /2\pi= 1.0$ Hz) and \ref{fig:freqDoub_signal} ($\Delta\omega/2\pi = 1.3$ Hz), the stationary evolution is periodic, attesting for the breakdown of the Maxwell-Bloch model.
Simulation parameters: $\omega_\textsc{fb} = G  m_0= 7.0$ Hz. The maser is triggered by a 10$^\circ$ initial flip angle.
}
\end{figure}

Results clearly show the transition from the Maxwell-Bloch regime, where the magnetization undergoes a series of monotonously decaying maser bursts and   $|m(t)|$ tends to a constant value, to a limit cycle behaviour for $(|m(t)|, m_z(t))$, where the stationary maser {\it envelope} evolves periodically, a situation that cannot be explained by Eqs. \ref{eq:MB_uv}.
It is striking that this transition occurs even for only a slight offset difference between the moments and whilst both lines are completely unresolved ($\Delta \omega/2\pi$ $=1$ Hz and $1.3$ Hz in the simulations, see Figs \ref{fig:linesep_sig_203}-\ref{fig:linesep_spec_203} and \ref{fig:freqDoub_signal}-\ref{fig:freqDoub_spectrum}). 
As expected, the associated spectrum consists of a Dirac comb with lines separated by $1/T$, where $T$ is the period of the maser pulses, and the envelope has the shape of a maser burst. 
Note that for $\Delta \omega /2\pi = 1.3$ Hz the observed pattern reproduces well the experiment of Fig. \ref{fig:methanol_single_mode}, and the structure of its Fourier spectrum is the sum of two  Dirac combs with a relative phase modulation (see the mathematical arguments in the Supplementary Material Eqs. \ref{eq_SI:periodicsignal_1} and \ref{eq_SI:spec_periodic_2}).

This toy model therefore points to the fact that  a structurally different kind of dynamics is expected even in the presence of unresolved, or partially resolved, multiplets.
Besides, the qualitative agreement between the methanol maser experiments (Figs. \ref{fig:Methanol_limitcycle_signal} -\ref{fig:methanol_driaccomb}) and the simulations reported in Figs.  \ref{fig:freqDoub_signal} and \ref{fig:freqDoub_spectrum} is noteworthy. 
The case of methanol, where the CH$_3$ resonance was completely unresolved is particularly interesting, as in a conventional treatment of NMR radiation feedback, it would be considered as a single magnetization characterized by an effective inhomogeneous transverse relaxation time  $T_2^*$.
However, this model obviously fails to account for the observed signal, which  shows that beyond a mere broadening, field inhomogeneity introduces structural complexity and requires to consider explicitly the effect of several magnetic moments coupled together through a common interaction with the detecting circuit.\cite{abergel_maser_2002, abergel_2009}
In this respect, our simulations thus show that this phenomenon is essentially contained in a simplistic toy model that is consistent with our observations and is able to account for a dynamics that is incompatible with the usual Maxwell-Bloch equations.

Finally, we discuss two limiting cases of the multiple component model that lead to approximate Maxwell-Bloch dynamics.
An obvious case is that of moments $m_k(t)$ with identical Larmor frequencies $\delta \omega_k=\delta\tilde\omega$ and relaxation rates $\gamma_{1,2}$.
 Then Eq. \ref{eq:fb_multilines} (or Eq. \ref{eq_SI:fb_multilines_compact}) reduces to the simple Maxwell-Bloch equations.
Alternatively, when all resonance frequency differences  $|\delta\omega_i - \delta\omega_k|$ are large, $\sin (\phi_i(t)-\phi_k(t) - \psi)$ and $\cos (\phi_i(t)-\phi_k(t) - \psi)$ are fast varying functions of the time as compared to $a_k(t)$and $m_{zk}(t)$ for all $k\ne i$.
Hence, for $\Delta t >> min(2 \pi/|\delta\omega_i - \delta\omega_k|)$,
the  averages $\bar a_i = \frac{1}{\Delta t} \int_{t_0}^{t_0+\Delta t} a(\tau) d\tau $, etc, obey the following relations:
\begin{eqnarray} \label{eq:MBE_approx}
 \nonumber  \dot{\bar a}_i(t) &=& -\gamma_{2i} \bar a_i - \gamma G m_{zi}(t) \bar a_i(t) \\
  \nonumber \dot{\bar m}_{zi}(t) &=& -\gamma_{1i} (\bar m_{zi} - m_{z0i})  + \gamma G \bar {a_i ^2}(t)  \cos( \psi)\\
 \dot{ \bar\phi}_i(t) &=& \delta \omega_i   + \gamma  G \bar m_{zi}(t)  \sin \psi,
\end{eqnarray}
which shows that in the case of large offset separations,  the moving average for individual   magnetizations ${\bf m}_i(t)$ also obeys Maxwell-Bloch equations, and each ${\bf m}_i(t)$ is decoupled from all other ${\bf m}_k(t)$.
Here, the set of moments ${\bf m}_i$ behaves as a multimode maser formed by the superposition of single mode masers obeying "on average" independent sets of Maxwell-Bloch equations.
In the former case, the feedback field is the sum of identical individual fields and acts on a degenerate line.
\begin{figure}
\centering
\subfigure{\label{fig:m_a_0p5Hz_spin_1_actual}
\begin{overpic}[scale=.22]{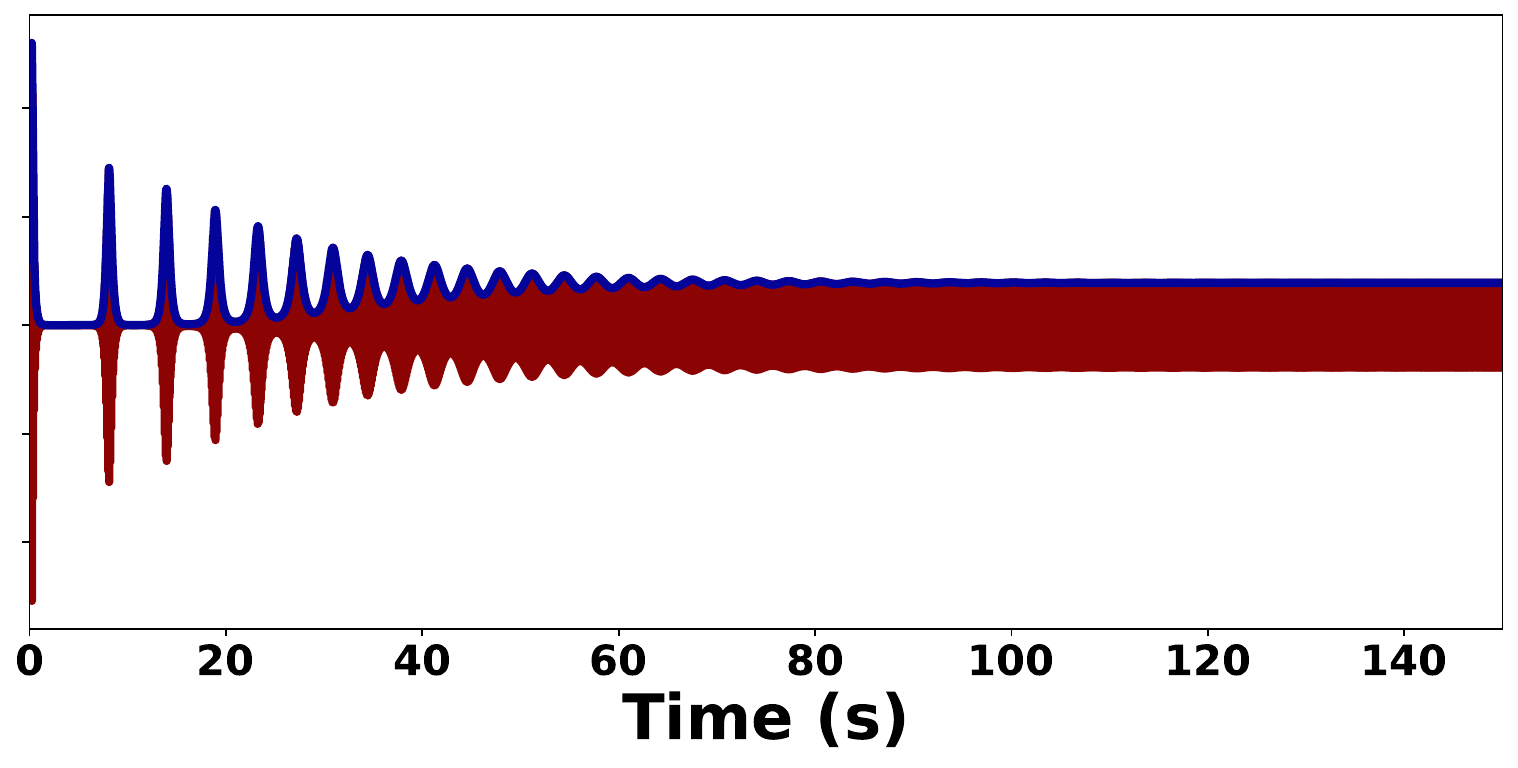}
\put(58,29){\includegraphics[scale=.09]{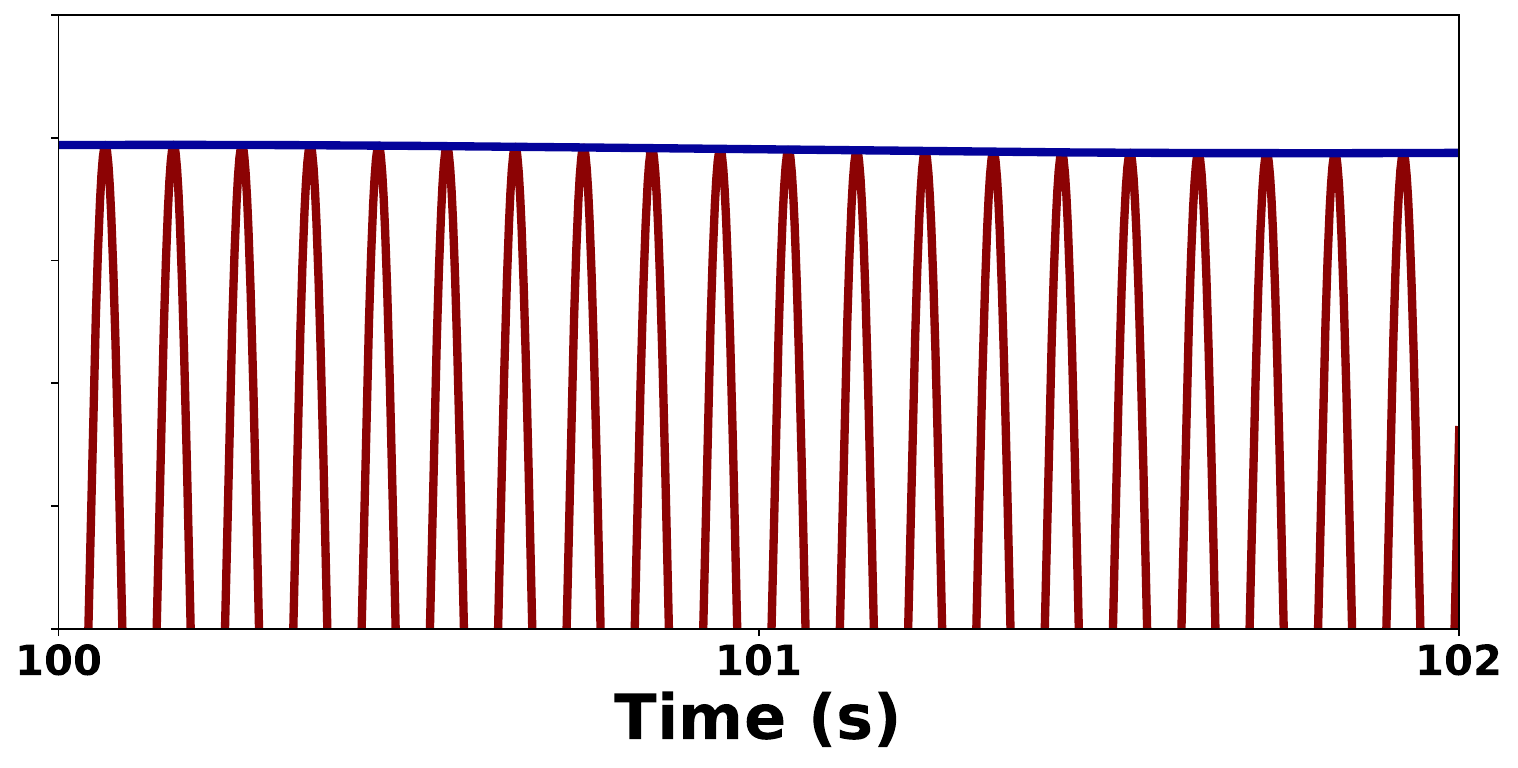} }      
\put(90,12){a)}
\end{overpic}
}
\subfigure{\label{fig:m_a_5Hz_2spin_signal}
\begin{overpic}[scale=.22]{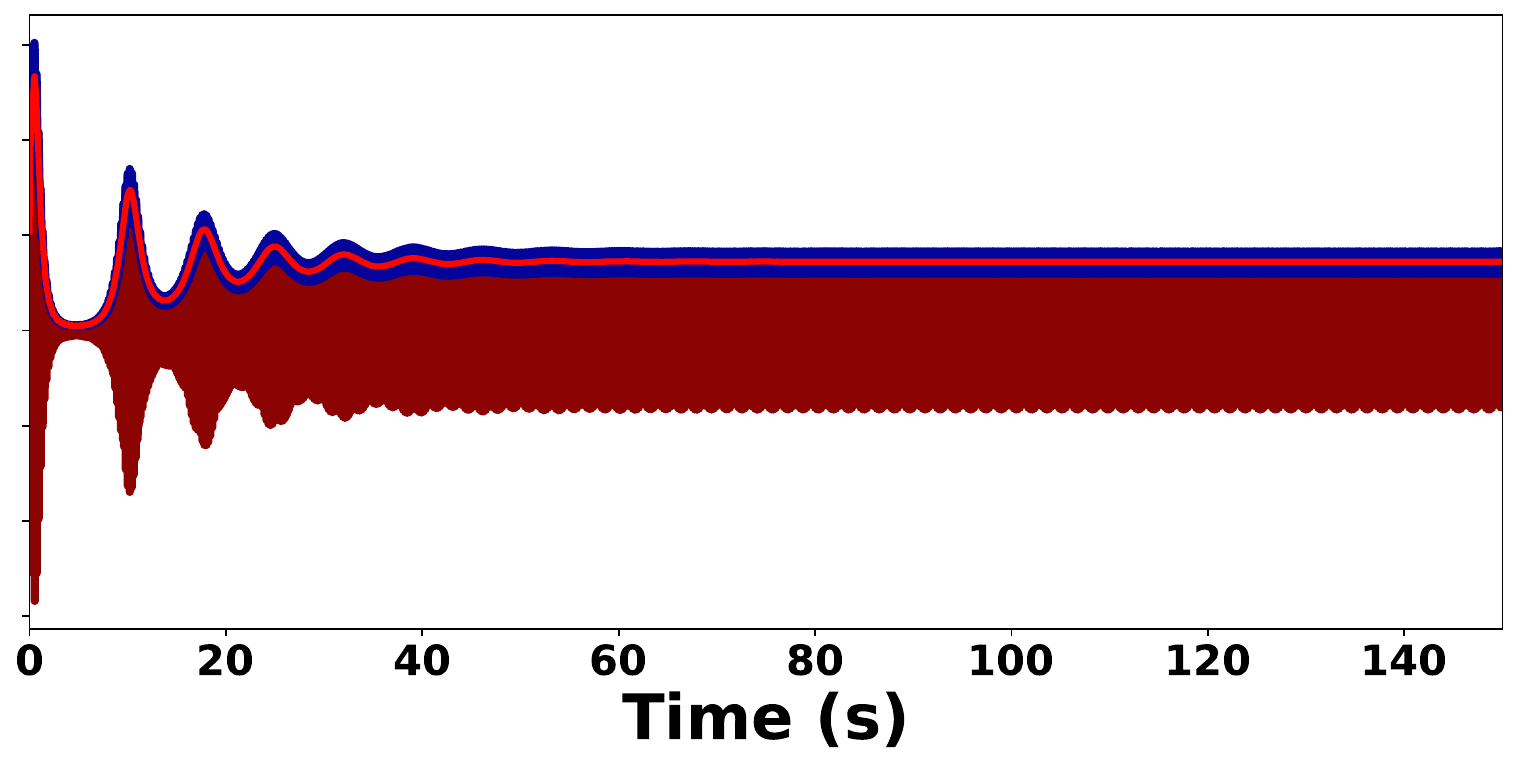}
\put(58,29){\includegraphics[scale=.09]{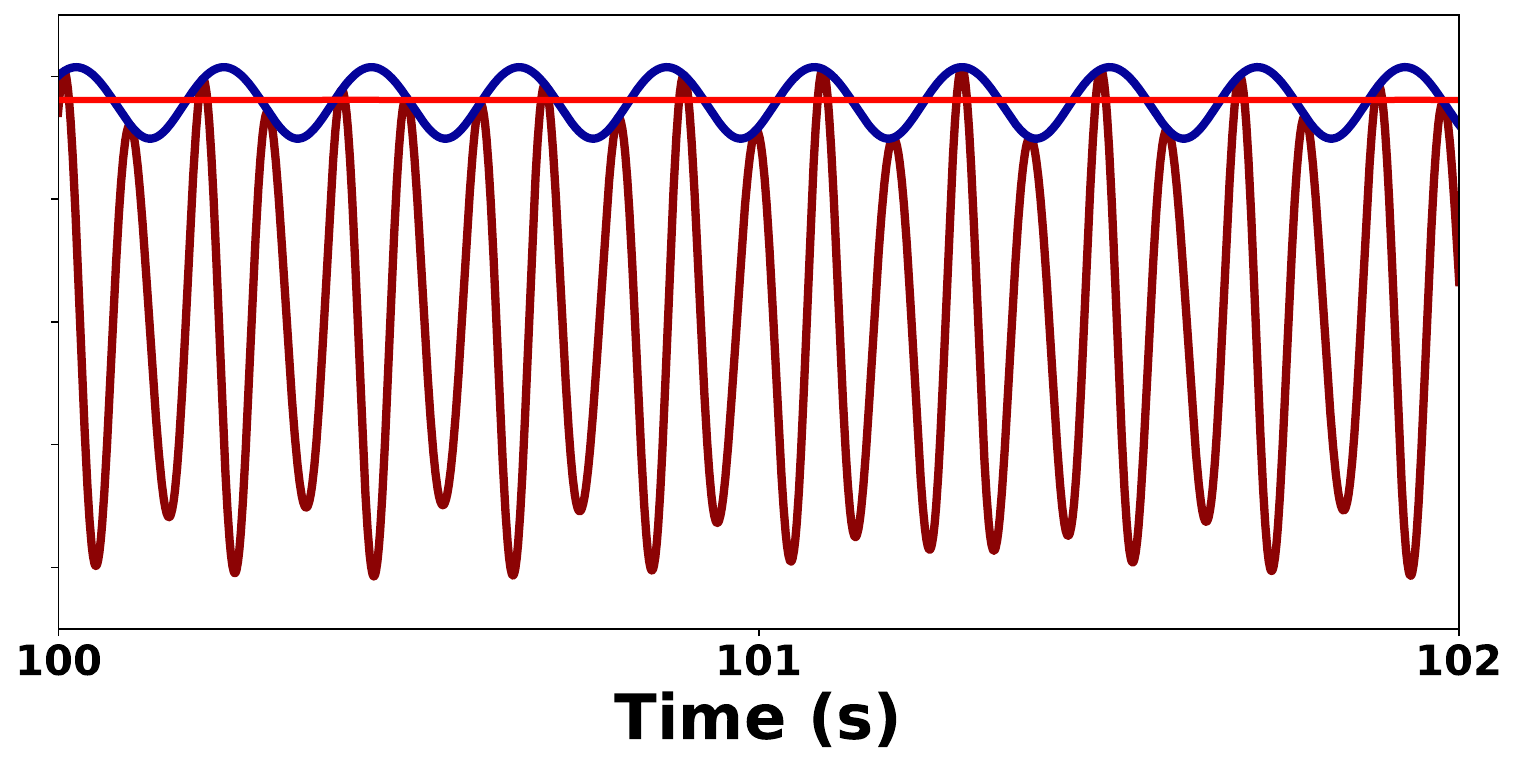}}    
\put(90,12){b)}
\end{overpic}
}
\subfigure{\label{fig:a1_5Hz_190Hz_2spin}
\begin{overpic}[scale=.22]{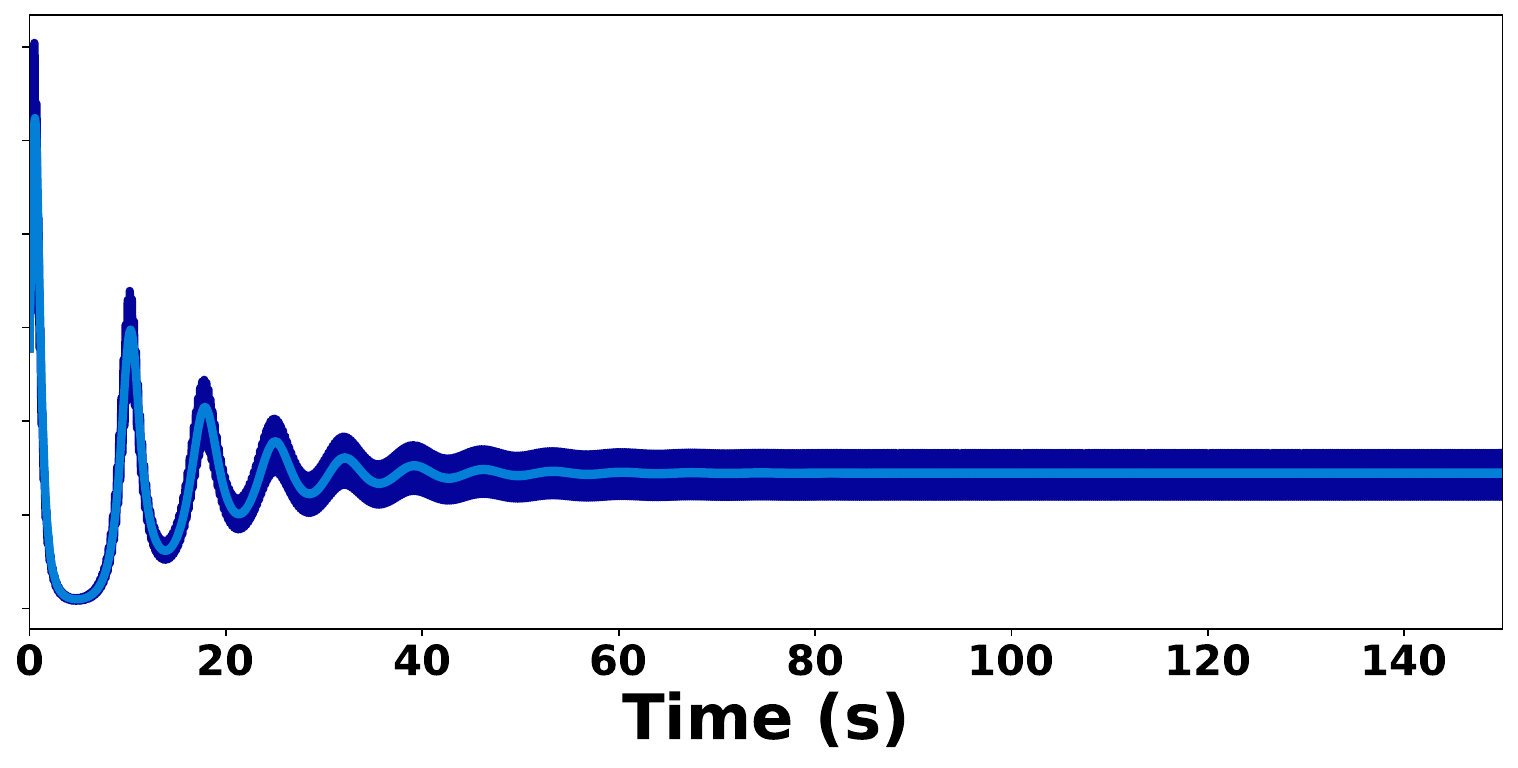}
    \put(58,29){\includegraphics[scale=.09]{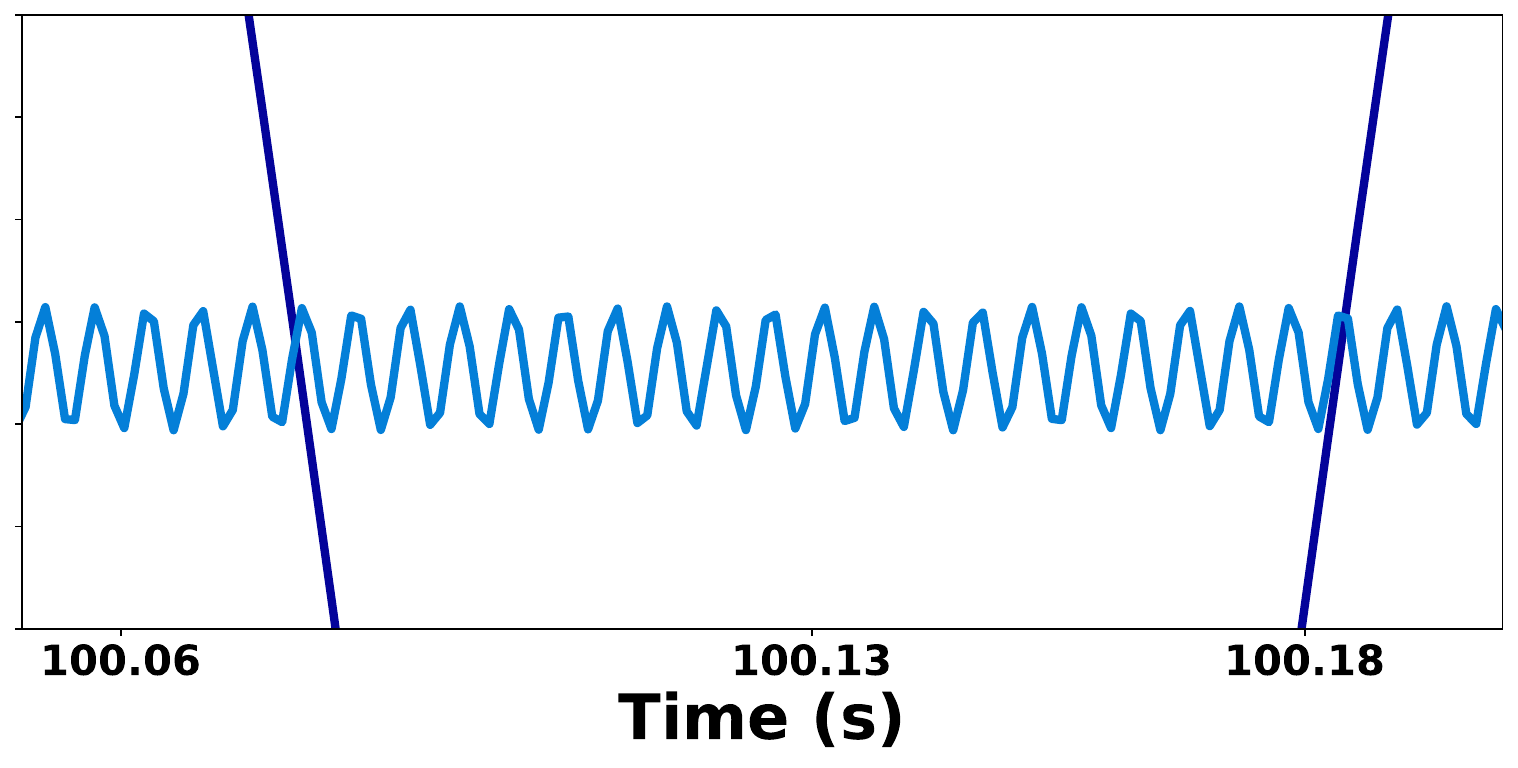}}
\put(90,12){c)}
\end{overpic}
}
\caption{\label{fig:lmiting_casesMaxwellBloch} Departure from the Maxwell-Bloch equations for the two-moment toy model of radiation feedback. 
Simulations are performed for moments with $\delta\omega_1/2\pi=10$ Hz and $\delta\omega_2/2\pi=10.5  (\ref{fig:m_a_0p5Hz_spin_1_actual}), 15  (\ref{fig:m_a_5Hz_2spin_signal}), \text{ and }200$ Hz.
Only the traces corresponding to the moment at  $\delta\omega_1/2\pi=10$ Hz is shown.
The transverse component $m_{1x}$ is depicted in brown.
The red trace  shows the envelope obtained from the  back calculated transverse magnetization from the filtered Fourier spectrum.
In \ref{fig:a1_5Hz_190Hz_2spin}, the dark blue and light blue traces correspond respectively  to the amplitude $a_1(t)$ for $\delta\omega_2/2\pi=15$ Hz and $\delta\omega_2/2\pi=200$ Hz.
} 
\end{figure}
These limiting cases are illustrated in Fig. \ref{fig:lmiting_casesMaxwellBloch} for two moments with different offset separations $\Delta\omega$.
For a small $\Delta\omega/2\pi = 0.5$ Hz (Figure \ref{fig:m_a_0p5Hz_spin_1_actual}), therefore unresolved  lines, the total transverse magnetization $m(t)$ exhibits the usual series of maser pulses with monotonous amplitude decay (brown trace) while its envelope (dark blue) decays to a  stationary value, in line with the usual Maxwell-Bloch equations.
When $\Delta\omega/2\pi = 5$ Hz (Figure \ref{fig:m_a_5Hz_2spin_signal}), ${\bf m}_1(t)$ still exhibits a series of decaying maser bursts (brown).
However, the number of these decreases before a steady state is reached, which is associated to a steady state oscillation of the  total transverse magnetization $m(t)$ (dark blue).  
Finally, the comparison of the decay profiles of the maser bursts for  large and small offset differences in Figure \ref{fig:a1_5Hz_190Hz_2spin} ($\Delta\omega/2\pi = 190$ Hz and $\Delta\omega/2\pi = 5$ Hz), shows that the amplitudes of $|{\bf m}_1(t)|$ are similar in both cases.
However, both traces show different steady state modulation of the amplitude $a_1(t)$ that is much slower and much larger for small $\Delta\omega$, where it closely matches the offset difference.
In contrast, for large $\Delta\omega$, this modulation is fast and hardly visible, thereby illustrating the Maxwell-Bloch behaviour of the average magnetization expected from Eqs. \ref{eq:MBE_approx}.
Interestingly, these oscillations are absent in the analysis of our experiments, as the strategy allowing to identify the  distinct masers is based on the filtering of the spectrum around individual resonances followed by an inverse Fourier transform.
This amounts to a convolution of the time signal with a time windowing function that averages out the fast variations due to the phase difference between the moments, which is similar in nature to the moving average process described in Eq. \ref{eq:MBE_approx}.
As a result, the time domain signal $m(t)$ actually reaches a constant value, rather than an oscillating steady state.

\paragraph*{\label{sec:ackn}{\bf Acknowledgements:}}
The authors acknowledge the support of the Agence Nationale pour la Recherche (ANR), Grant ANR-22-CE29-0006-01–DynNonlinPol.
\bibliography{biblio}
\makeatletter\@input{letter_SI_aux.tex}\makeatother
\end{document}



\title{ Supplementary information - 
Multi-mode masers of thermally polarized nuclear spins in solution NMR}

\author{Vineeth Francis Thalakottoor Jose Chacko}
\email{vineeth.thalakottoor@ens.psl.eu}
\affiliation{%
 Laboratoire des Biomolécules, LBM, Département de Chimie, Ecole Normale Supérieure, PSL University, Sorbonne Université, CNRS, 75005 Paris, France
}%
\author{Alain Louis-Joseph}
\email{alain.louis-joseph@polytechnique.edu}
\affiliation{
 Laboratoire de Physique de la Matière Condensée, UMR 7643, CNRS, École Polytechnique , IPP 91120 Palaiseau, France
}%
\author{Daniel Abergel}%
\email{daniel.abergel@ens.psl.eu}
\affiliation{%
 Laboratoire des Biomolécules, LBM, Département de Chimie, Ecole Normale Supérieure, PSL University, Sorbonne Université, CNRS, 75005 Paris, France
}%
\date{\today}
\maketitle

\section*{Stability analysis of the Maxwell-Bloch equations}
The dynamics of the magnetization is examined by first determining the fixed points (the stationary solution of $\frac{d{\bf m}}{dt}=0$) and then analyzing the stability of the linearized system in their vicinity.\cite{holmes}
These fixed points have been found in the case where $\omega_1=0$, and it was shown that for certain values of the relaxation and feedback field  parameters, the $(m^2,m_z)$ system admits an out of thermal equilibrium stable fixed point.\cite{abergel_maser_2002, ottochaos_2023} 
that can be a stable focus, meaning that the trajectory of $(m^2(t),m_z(t))$  spirals inwards towards its asymptotic value (see Eqs \ref{eq_SI:F2}-\ref{eq_SI:vpF1} in the SI).
This motion exactly represents a series of maser pulses with a monotonous decrease of their intensities.
This discussion\cite{abergel_maser_2002,Weber2019} is summarized below.

Denoting  $m_t=m_x+i m_y = \sqrt{u}e^{i\phi(t)}$, the MB equations become:
\begin{equation}\label{eq:MB_uv}
\left \lbrace
\begin{array}{l}
\displaystyle \dot u(t) =  2(\lambda m_z(t)\sin \psi  - \gamma_2) u(t)\\
\displaystyle \dot m_z(t) = -\lambda \sin \psi u(t) -\gamma_1 (m_z(t) - m_0) \\
\displaystyle  \dot \phi(t) = -\delta + \lambda \cos \psi\; z
\end{array}
\right .
\end{equation}
The fixed points are thus $F_1=(0,m_0)^t$ and $F_2= (-\frac{\gamma_1}{\lambda \sin \psi} \left [ \frac{\gamma_2}{\lambda \sin \psi} - m_0\right ], \frac{\gamma_2}{\lambda \sin \psi})^t$, and their stability dictates the dynamics of the system.
An analysis shows that, depending on the parameters,  one of $F_1$ and $F_2$ is stable and the other is unstable.
Thus, when the product $m_0 \times \sin \psi  < 0$, $F_2$ is unstable and $F_1$ is stable. 
This corresponds to the usual radiation damping case, where the radiation feedback field from the probe drives the magnetization towards $+z$.
In contrast, for $m_0 \times \sin \psi  > 0$, two situations occur and $F_2$ is stable
if the condition $ \lambda \sin \psi m_0  -\gamma_2   >  0 $ is fulfilled.
When in addition $\Delta = \gamma_z(\gamma_z+8\gamma_2  -  8 \lambda m_0 \sin \psi)$, then $F_2$ is a focus.
This describes the evolution of a sustained maser  towards a stationary magnetization that precesses about the $z $ axis on a cone of semi-angle $\alpha$ such that $\tan\alpha = \frac{\lambda \sin \psi \sqrt{u}}{\gamma_2}$.
The physical interpretation of the necessary condition  $m_0 \times \sin \psi  > 0$ for such dynamics to occur is the existence of two competing processes originating from the feedback field and longitudinal relaxation.
\subsection{Fixed points of the linearized Maxwell-Bloch equations}
Linearization at:
\begin{equation}\label{eq_SI:F2}
 F_2 = (-\frac{\gamma_z}{\lambda \sin \psi} \left [ \frac{\gamma_2}{\lambda \sin \psi} - m_0\right ], \frac{\gamma_2}{\lambda \sin \psi},m_0)
\end{equation}
\begin{eqnarray} \label{eq_SI:lin_F2}
\nonumber \left [ \begin {array}{c} \dot U\\\dot Z \\ \dot W \end{array}\right ] &=& 
\left[\begin {array}{ccc}
    0 & 2\lambda\sin \psi\; u^{st}&0\\
    -\lambda \sin \psi &-\gamma_z& \gamma_z \\
    0 & 0 & -\gamma_{st}
  \end{array}\right ]
\left[\begin {array}{c}  U\\ Z\\ W\end{array}\right ]  \\
 &+& \left [ \begin {array}{c}  -2\lambda \sin \psi  ZU\\ 0 \\0\end{array}\right ]
\end{eqnarray}
The eigenvalues are:
\begin{eqnarray}\label{eq_SI:vpF2}
\nonumber x_0&=& -\gamma_{st}\\
  x_{\pm} &=& \frac{-\gamma_z \pm \sqrt{\Delta}}{2} \textrm{,  }\Delta = \gamma_z(\gamma_z+8\gamma_2  -  8 \lambda m_0 \sin \psi)
\end{eqnarray}
Now, if  the condition $m_0 \times \sin \psi  < 0$ then one has: $\Delta > \text{ and }x_+ >0, \; x_- <0$

Linearization at:
\begin{equation}\label{eq_SI:F1}
  F_1 = (0,m_0, m_0) 
\end{equation}
\begin{eqnarray}\label{eq_SI:lin_F1}
\nonumber \left [ \begin {array}{c} \dot U\\\dot Z \\ \dot W \end{array}\right ] &=& 
\left[\begin {array}{ccc}
    2( \lambda\sin \psi - \gamma_2)&    0 &0\\
    -\lambda \sin \psi &-\gamma_z& \gamma_z \\
    0 & 0 & -\gamma_{st}
  \end{array}\right ]
\left[\begin {array}{c}  U\\ Z\\ W\end{array}\right ]  \\
 &+& \left [ \begin {array}{c}  -2\lambda \sin \psi  ZU\\ 0 \\0\end{array}\right ]
\end{eqnarray}
\begin{eqnarray}\label{eq_SI:vpF1}
\nonumber x_0&=& -\gamma_{st}\\
\nonumber  x_{1} &=& -\gamma_z \\
 x_2 &=& 2(\lambda \sin \psi m_0 - \gamma_2)
\end{eqnarray}
\subsection{Fixed point stability }
\begin{itemize}
    \item Fixed point stability in the case $m_0 \times \sin \psi  < 0$. This corresponds to the radiation damping case.  In this case, one has: 
    \begin{itemize}
        \item If $\Delta > \text{ and }x_+ >0, \; x_- <0$
    \end{itemize}
    \item Fixed point stability in the case $m_0 \times \sin \psi  > 0$
    \begin{itemize}
        \item if $  \lambda \sin \psi m_0  -\gamma_2   >  0$ then $x_2>0$ and $F_1$ is unstable, and $F_2$ is stable.
        \begin{itemize}
            \item If in addition $\Delta < 0$, then $F_2$ is a stable focus.
        \end{itemize}
        \item If $  0<\lambda \sin \psi m_0  <\gamma_2 $, then $x_2< 0$ and  $F_1$ is stable, and $F_2$ is unstable.
    \end{itemize}

\end{itemize}
\section*{A limit cases of Eq.\ref{eq:fb_multilines}}
The nonlinear dynamical system of Equations \ref{eq:fb_multilines} can  essentially be investigated by  numerical simulations.
To this aim equations can be reformulated, as proposed in ref.\cite{appelt_jmr_2021} (or in Equation \ref{eq:MB_uv} above\cite{abergel_maser_2002}), by introducing the amplitudes and phases $a_k(t)$ and $\phi_k(t)$ of each moment $m_k(t)$: $m_k(t) = a_k(t) e^{i\phi_k(t)}$.
The evolution of one of the moments ${\bf m}_i$ due to the effect of the feedback field is thus:
\begin{eqnarray} \label{eq_SI:fb_multilines_compact}
\nonumber     \dot m_{i}&=& -(i \delta \omega_i + \gamma_{2i}) m_i  - i \omega_1 m_{zi}  e^{i\psi_1} +i\gamma G m_{zi} e^{-i\psi}\sum_k a_k  e^{i\phi_k }\\
    \dot m_{zi}&=&  - \omega_1 {\cal R}e(i e^{-i\psi_1}m_i  )  -\gamma_{1i} (m_{zi} - m_{z0i}) -  \gamma G a_i  \sum_k a_k  \sin(\phi_i  -\phi_k  + \psi)
\end{eqnarray}
and the differential equations for the amplitudes and phases are thus:
\begin{eqnarray}\label{eq_SI:MB_multiline_ampl_phase_mz}
\nonumber    \dot a_i &=&-\gamma_{2i} a_i +\omega_1 m_{zi}  \sin(\phi_i + \psi_1) + \gamma G m_{zi} \sum_k a_k  \sin(\phi_i - \phi_k +\psi)\\
    \dot \phi_i &=& \delta \omega_i + \frac{\omega_1}{a_i}m_{zi} \cos(\phi_i + \psi_1)-\frac{\gamma G m_{zi} }{a_i } \sum_k a_k  \cos(\phi_i - \phi_k +\psi)\\
 \nonumber    \dot m_{zi}&=& \omega_1 a_i \sin(\phi_i + \psi_1)  -\gamma_{1i} (m_{zi} - m_{z0i})  - \gamma G a_i  \sum_k a_k  \sin(\phi_i  -\phi_k  + \psi)
\end{eqnarray}

\begin{figure}[h!]
\centering
\subfigure[]{\label{fig_SI:Methanol_REF}\includegraphics[width=.45\textwidth]{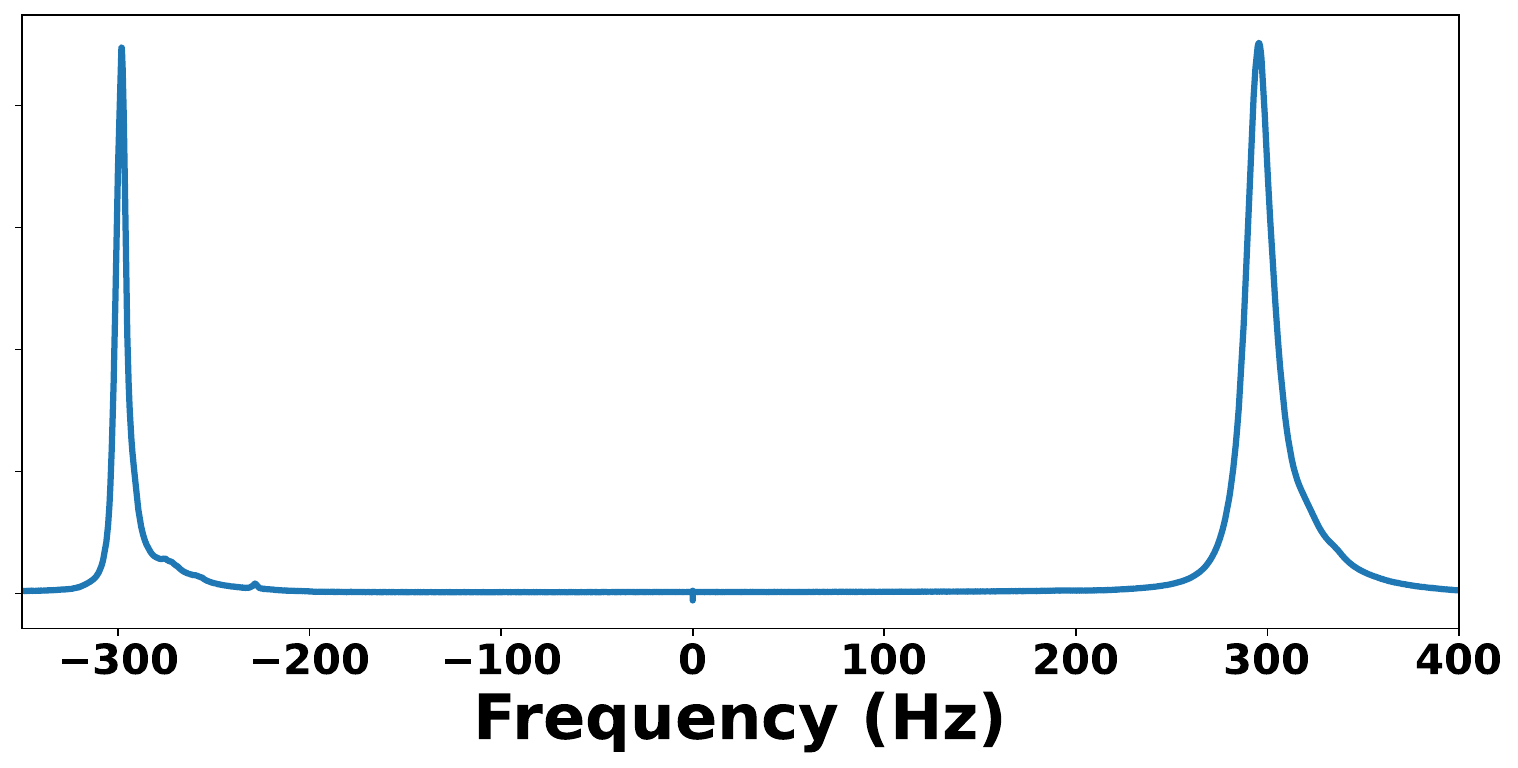}}
\subfigure[]{\label{fig_SI:Ethanol_REF}\includegraphics[width=.45\textwidth]{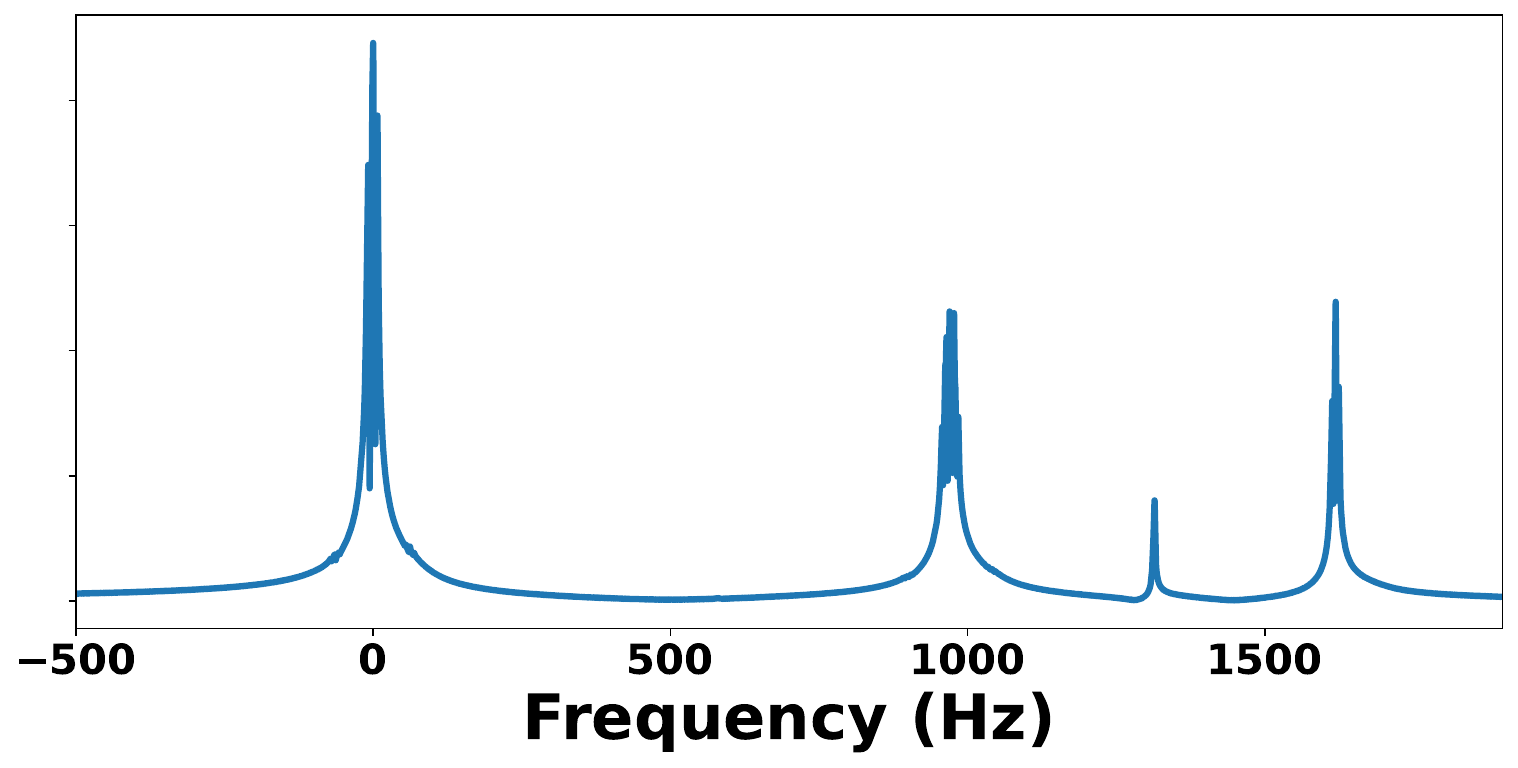}} 
\caption{\label{fig_SI:Methanol_ethanol_Ref} Reference Spectra of  Methanol (\ref{fig_SI:Methanol_REF}) and Ethanol(\ref{fig_SI:Ethanol_REF}). Frequencies are labelled with the LO as the reference. Spectra were acquired at 9.4 T. For the methanol, the lines are unresolved due to the addition of CuSO$_4$ in the solution.
} 
\end{figure}

In general, equations \ref{eq_SI:MB_multiline_ampl_phase_mz} can only be studied by numerical simulations.
Nevertheless, qualitative results can be obtained in two limiting cases, when $\omega_1 = 0$.
First, assume that 
all the moments $m_k(t)$ have nearly identical Larmor frequencies $\delta \omega_k=\delta\tilde\omega$, and relaxation rates $\gamma_{1,2}$.
Then, from Equation \ref{eq_SI:fb_multilines_compact}, one has:
\begin{eqnarray}\label{eq:fb_multilines_monofrequency_mt}
    \dot m(t)&=&  -i \delta \tilde \omega \sum_i m_i(t) -\gamma_{2}\sum_i m_i(t)  + i\gamma G  e^{i\psi} \sum_i m_{zi}(t) \sum_k m_k(t) \\
      &=& -(i \delta \tilde \omega +\gamma_{2} ) m(t)  + i\gamma G  e^{i\psi} m_{z}(t) m(t) 
\end{eqnarray}
In the case where all pairwise resonance frequency differences  $|\delta\omega_i - \delta\omega_k|$ are large, and so are the differences $|\phi_i(t)-\phi_k(t)|$, then  the functions $\sin (\phi_i(t)-\phi_k(t) - \psi)$ and $\cos (\phi_i(t)-\phi_k(t) - \psi)$ are fast varying functions of the time as compared to the  amplitudes $a_k(t)$and $m_{zk}(t)$ components of the moments, for all $k\ne i$.
Therefore, the  average $\bar a_i = \displaystyle \frac{1}{\Delta t} \int_{t_0}^{t_0+\Delta t} A(\tau) d\tau $ obeys the following relations:
\begin{eqnarray}\label{eq:amplitude_ave}
\nonumber  \Delta t \bar a_i &=&\int_{t_0}^{t} -\gamma_{2i} a_i - \gamma G m_{zi} \sum_k a_k  \cos(\phi_i - \phi_k -\psi) d\tau\\
  &\approx& - \int_{t_0}^{t} \left [\gamma_{2i} a_i   + \gamma G \cos\psi\, m_{zi}(\tau) a_i(\tau)\right ]  d\tau
  - \gamma G  \sum_{k\ne i}  m_{zi} (t) a_k(t) \int_{t_0}^{t} \cos(\phi_i - \phi_k -\psi) d\tau\\
\nonumber &\approx&  - \int_{t_0}^{t} \left [\gamma_{2i} a_i   + \gamma G  m_{zi}(\tau) a_i(\tau)\right ]  d\tau
\end{eqnarray}
Similarly, one obtains for the moving averages $\bar \phi_i$ and $\bar m_{zi} $ the following equations:
\begin{eqnarray}\label{eq:amplitude_ave_phi_mz}
  \Delta t \bar\phi_i(t) &\approx&  \int_{t_0}^{t} \left [\delta \omega_i   +  \gamma G m_{zi}(t)  \sin \psi\right ] d\tau \\
  \Delta t \bar m_{zi}(t)&\approx& \int_{t_0}^{t}  \left [-\gamma_{1i} (m_{zi} - m_{z0i})  + \gamma G a_i ^2(\tau)  \cos( \psi)\right ]\,  d\tau
\end{eqnarray}
Thus, the dynamics of the component ${\bf m}_i(t)$ obeys the approximate evolution equations:
\begin{eqnarray} \label{eq_SI:MBE_approx}
  \dot{\bar a}_i(t) &=&   -\gamma_{2i} \bar a_i   - \gamma G  m_{zi}(t) \bar a_i(t) \\
  \dot{\bar m}_{zi}(t) &=& -\gamma_{1i} (\bar m_{zi} - m_{z0i})  + \gamma G \bar {a_i ^2}(t)  \cos( \psi)\\
 \dot{ \bar\phi}_i(t) &=& \delta \omega_i   + \gamma  G \bar m_{zi}(t)  \sin \psi
\end{eqnarray}
This therefore shows that in this case of large offset separations,  the moving average for the  magnetization ${\bf m}_i(t)$ again obeys Maxwell-Bloch equations, and each ${\bf m}_i(t)$ is decoupled from all other ${\bf m}_k(t)$.
\newpage
\section*{Figure \ref{fig_SI:Methanol_ethanol_Ref}}
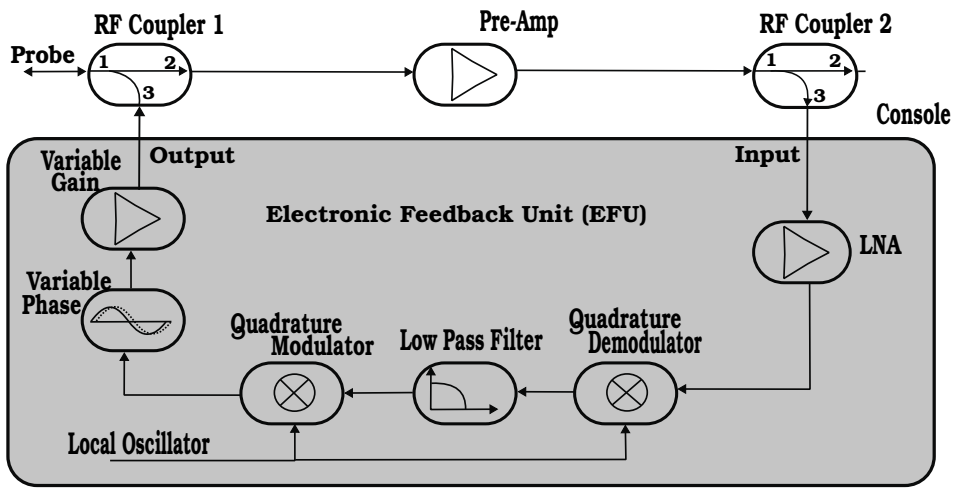
\begin{figure}[h]
\begin{overpic}[width=.7\textwidth]{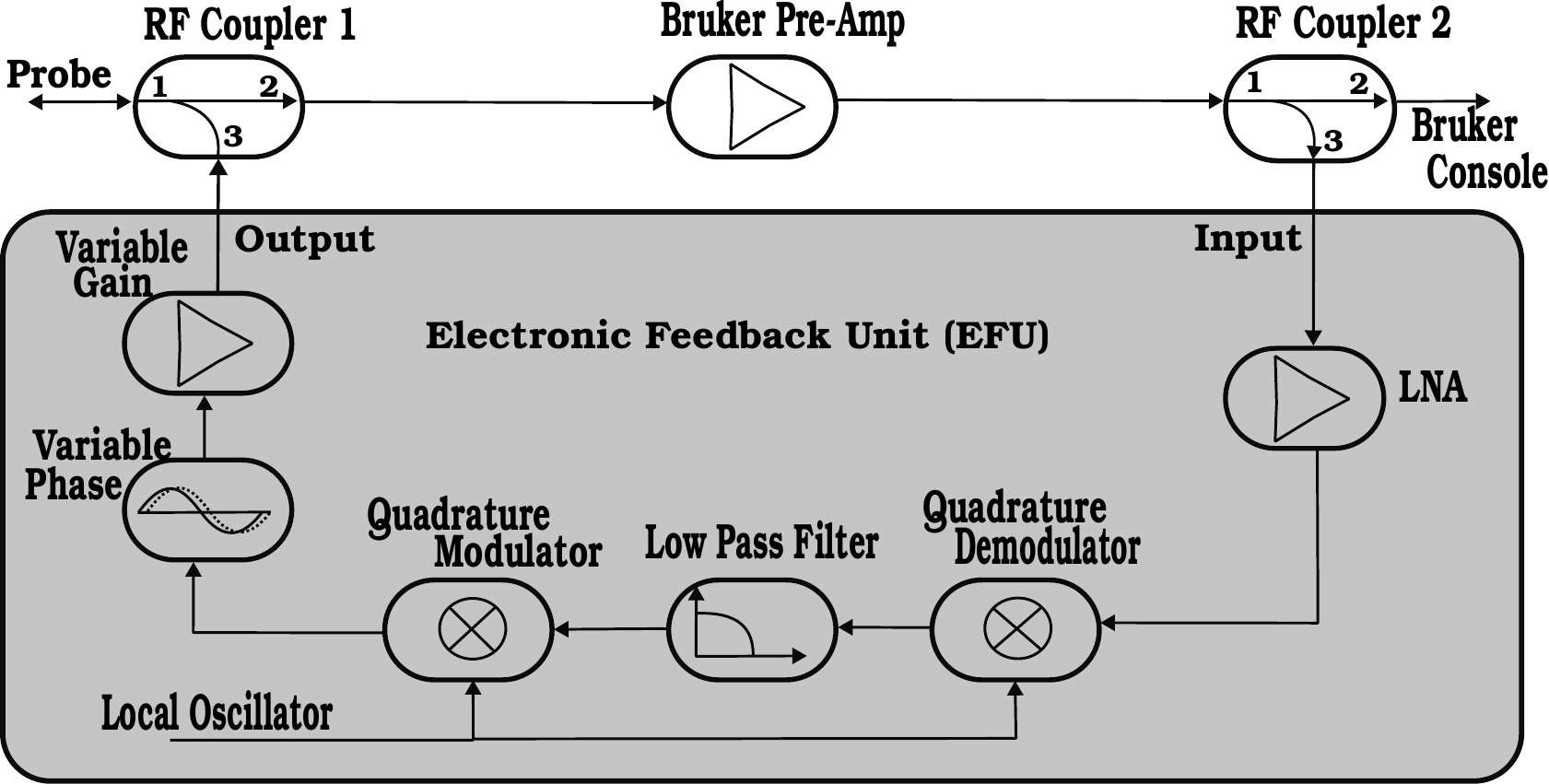}
     \put(42,48){\tikz \draw[fill=white,color=white,thick] (0,0) rectangle (1,0.4);}
     \put(91,41){\tikz \draw[fill=white,color=white,thick] (0,0) rectangle (0.9,0.5);}
\end{overpic}
\caption{\label{fig:RDCU}\small Electronic Feedback Control Unit (eFCU)- The signal from the probe is picked up at the output of the preamplifier through the directional coupler "2" and amplified by a  low-noise amplifier (+20 dB gain).
Demodulation at 400 MHz (local oscillator - LO)  was used  for sake of filtering, and picked up from the $^1$H power amplifier of the spectrometer. Its power set to a 14.7 dBm.
The re-modulated signal is then adjusted in phase and gain before being re-injected into the probe through the rf coupler 1.
Gain adjustment was performed by a +20 dB low noise amplifier and a set of variable attenuators in the range 0 to $111$ dB.}
\end{figure}

\newpage
\section*{Figure \ref{fig:RDCU} - Instrumentation description}
A fraction of the induction signal is picked up at the output of the preamplifier through a directional coupler and amplified through a low-noise amplifier.
This signal is then demodulated using a 400 MHz local oscillator (LO) reference generated from the $^1$H power amplifier of the spectrometer.
The resulting in-phase and quadrature demodulated signals are fed into a low-pass filter of bandwidth 100 kHz and re-modulated at the proton Larmor frequency.
After re-modulation the radiofrequency signal is phase adjusted by a voltage-controlled phase modulator, and the gain of the eFCU output signal was controlled using a second LNA (+20 dB) and a set of variable attenuators in the range 0 to $111$ dB (with a minimum attenuation step of 0.1 dB).
The LO was switched on throughout acquisition.
Finally, the phase- and gain-adjusted signal is fed back into the probe via an additional directional coupler.
The rf leakage of the LO from the mixers of the eFCU is first minimized ($\lesssim  2$ mV)  by adjusting the offset voltage at the input of the low-pass filter.

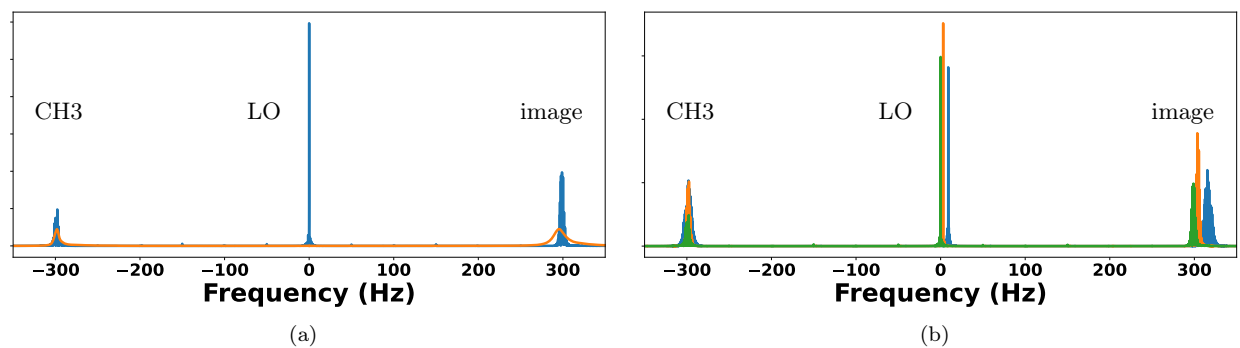
\begin{figure}[h!]
\centering
\subfigure[]{\label{fig_SI:Methanol_position_LO_Center}
\begin{overpic}[width=.45\textwidth]{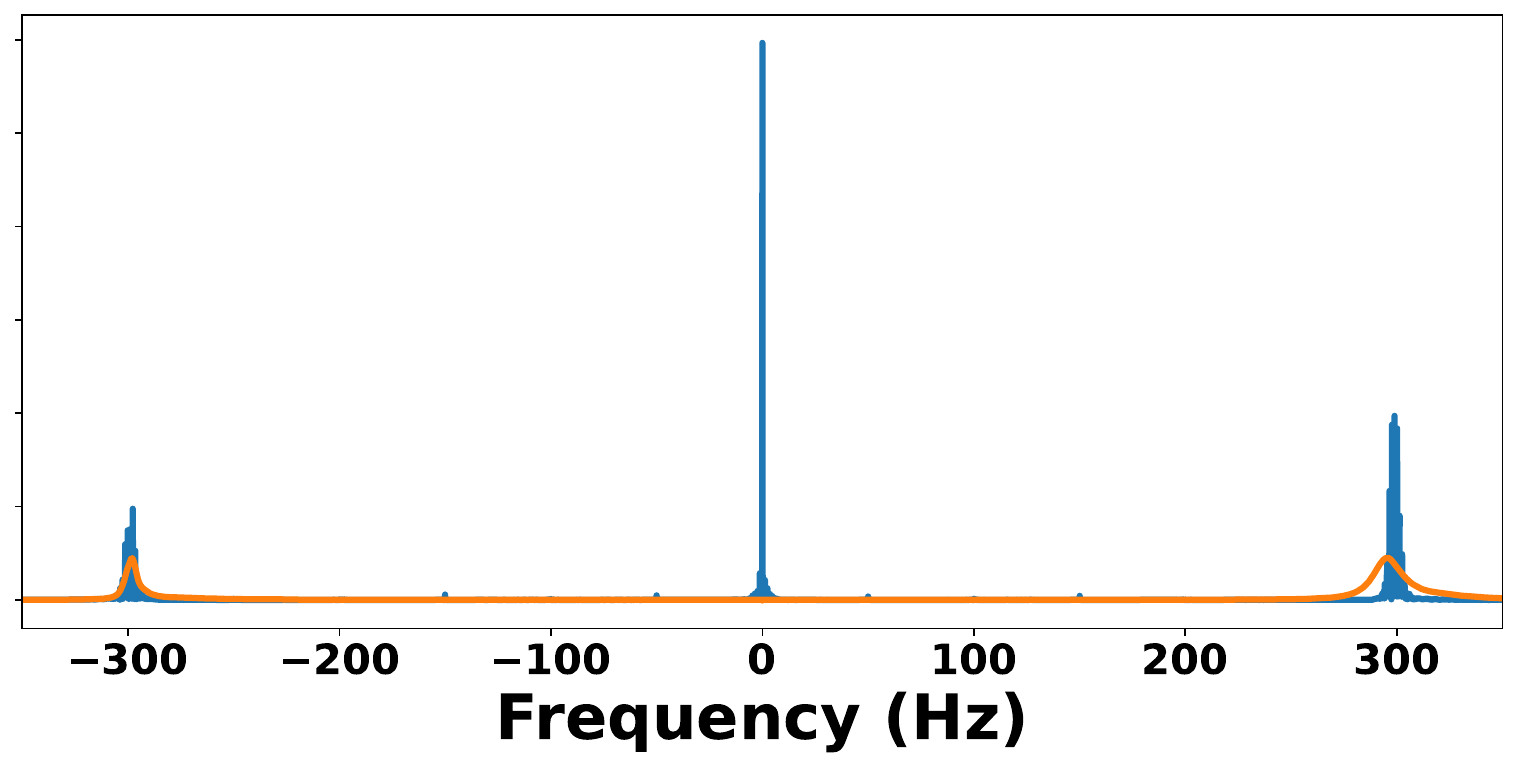}
    \put(40,32){LO }
    \put(5,32){CH3}
    \put(85,32){image}
\end{overpic}

}
\subfigure[]{\label{fig_SI:Methanol_position_LO_multiple}
\begin{overpic}[width=.45\textwidth]{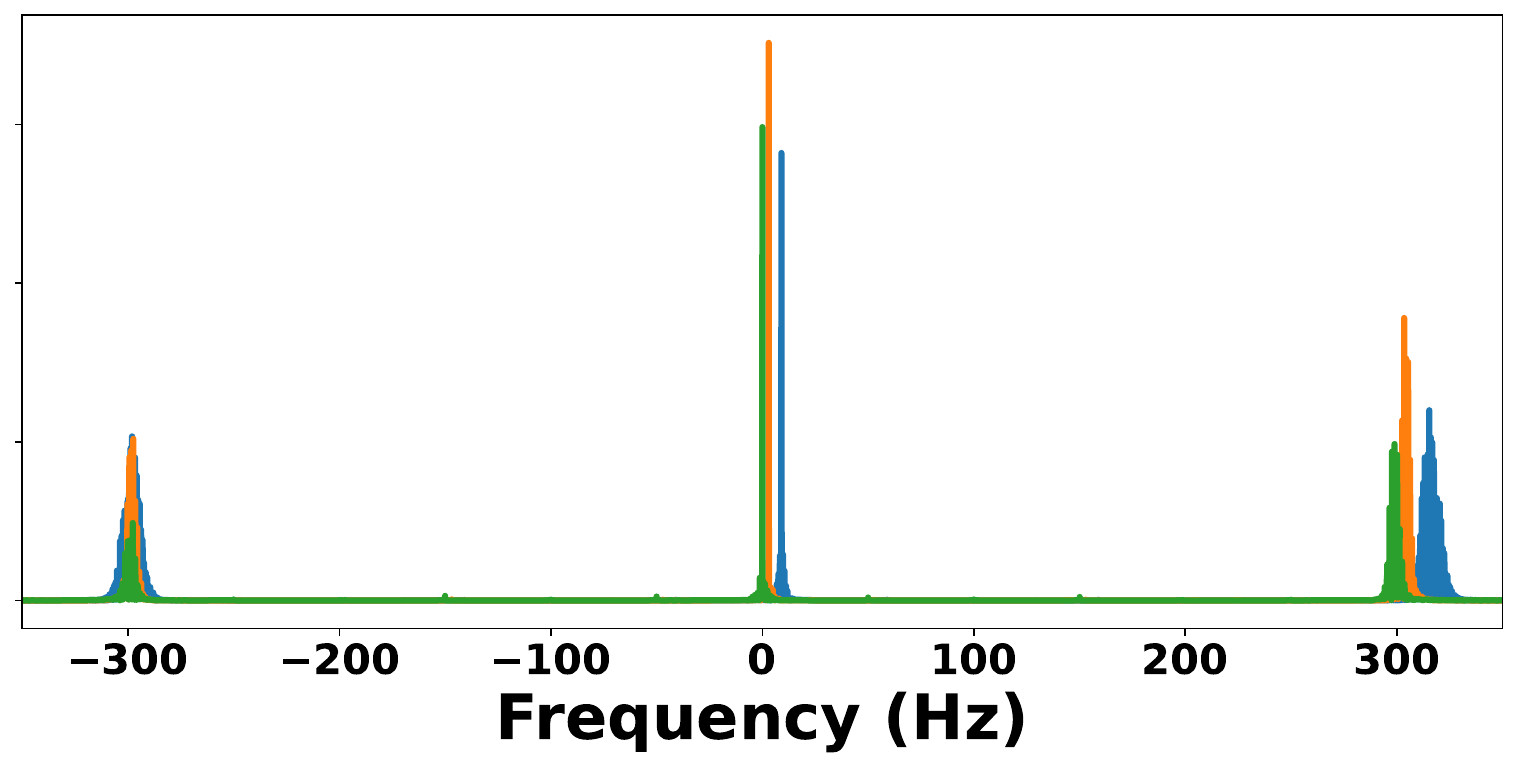}
    \put(40,32){LO }
    \put(5,32){CH3}
    \put(85,32){image}
\end{overpic}
} 

\caption{\label{fig_SI:Methanol_position_LO} Effect of modifications of the modulation/demodulation reference frequency (LO) on the methanol resonance line in maser experiments - Spectra of a methanol maser from CH3 for different LO positions. In \ref{fig_SI:Methanol_position_LO_Center}: a maser spectrum (blue) is superimposed with a reference FID (orange); the zero frequency corresponds to he LO position for the spectrum on the left. In \ref{fig_SI:Methanol_position_LO_multiple}: Changing the LO frequency only changes the image location whilst the CH3 resonance line remains fixed. The carrier frequency is the same for all experiments.
} 
\end{figure}

Mirror images that are symmetric with respect to the demodulation frequency are caused by the low pass filtering of the signal that requires a demodulation/remodulation stage of the audio signal in the eFCU.
Demodulation of the input signal $\nu_0+\nu$, where $\nu_0$ is the LO radiofrequency  and $\nu=\delta \omega/2\pi$ is the audiofrequency of the spin, yields signals at $2\nu_0+\nu$ and $\nu$. 
The former is filtered out by the eFCU low-pass filter, and the latter gives signals at $\nu_0\pm\nu$  after re-modulation with the same LO frequency $\nu_0$.
The small and constant rf leakage yields a spike in the spectrum at the LO frequency.
These features are illustrated in Fig. \ref{fig_SI:Methanol_position_LO}.
\newpage\section*{Figure \ref{fig_SI:Methanol_position_LO} }
\begin{figure}[h!]
\centering
\subfigure[]{\label{fig_SI:B1_measure_nutation_methaol}\includegraphics[width=.45\textwidth]{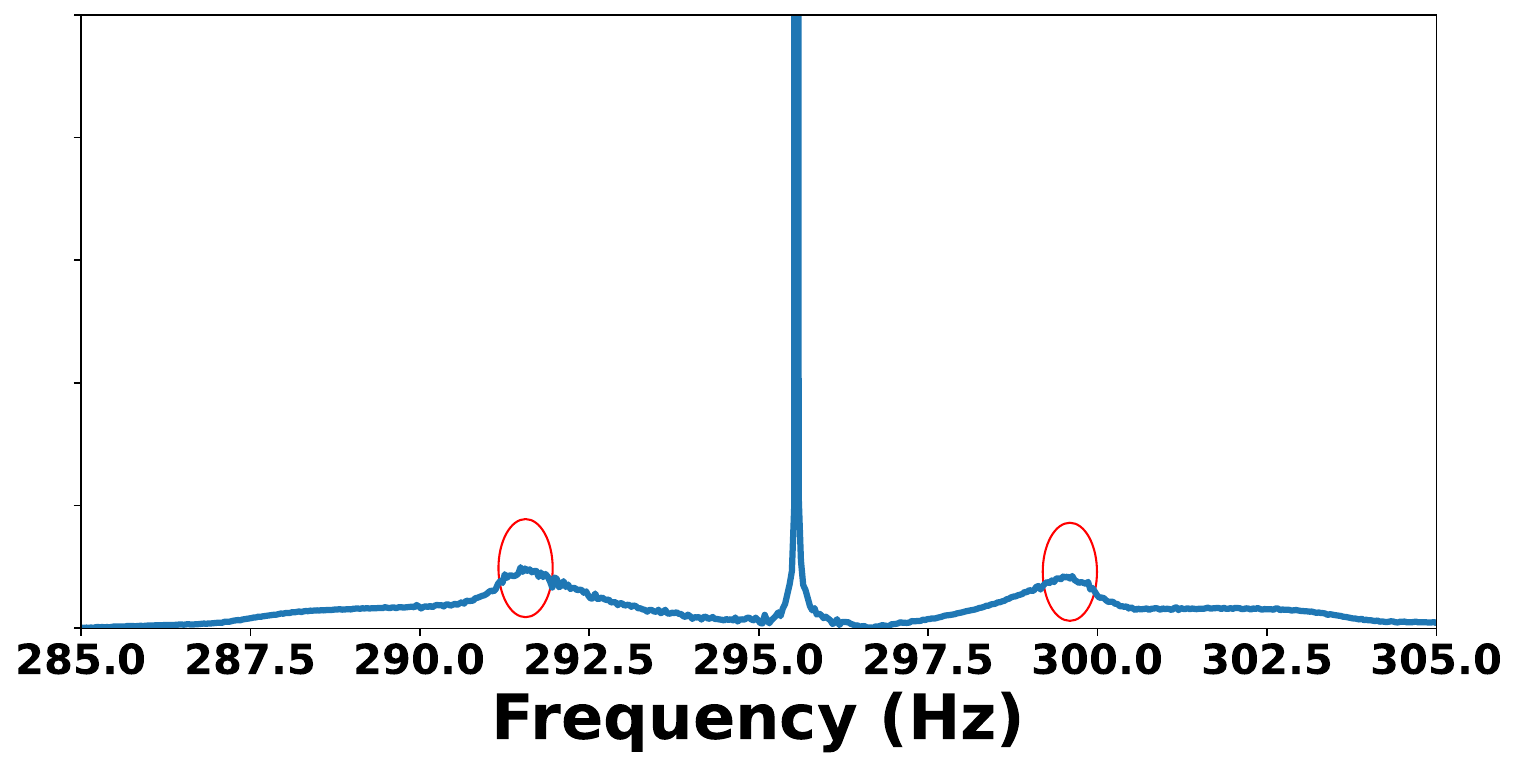}}
\subfigure[]{\label{fig_SI:B1_measure_nutation_methaol_sim}\includegraphics[width=.45\textwidth]{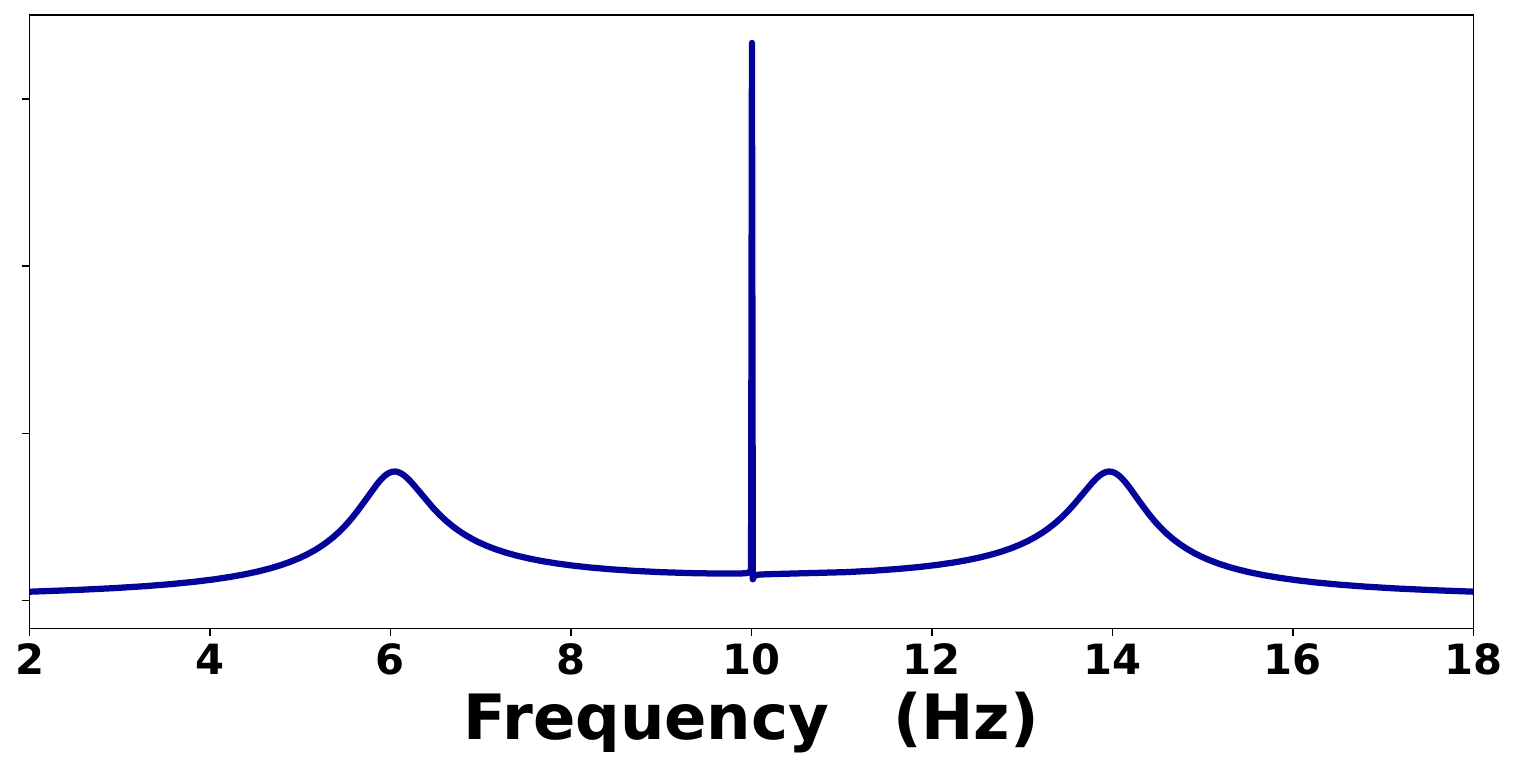}} 
\\
\subfigure[]{\label{fig_SI:B1_measure_nutation_ethaol}\includegraphics[width=.45\textwidth]{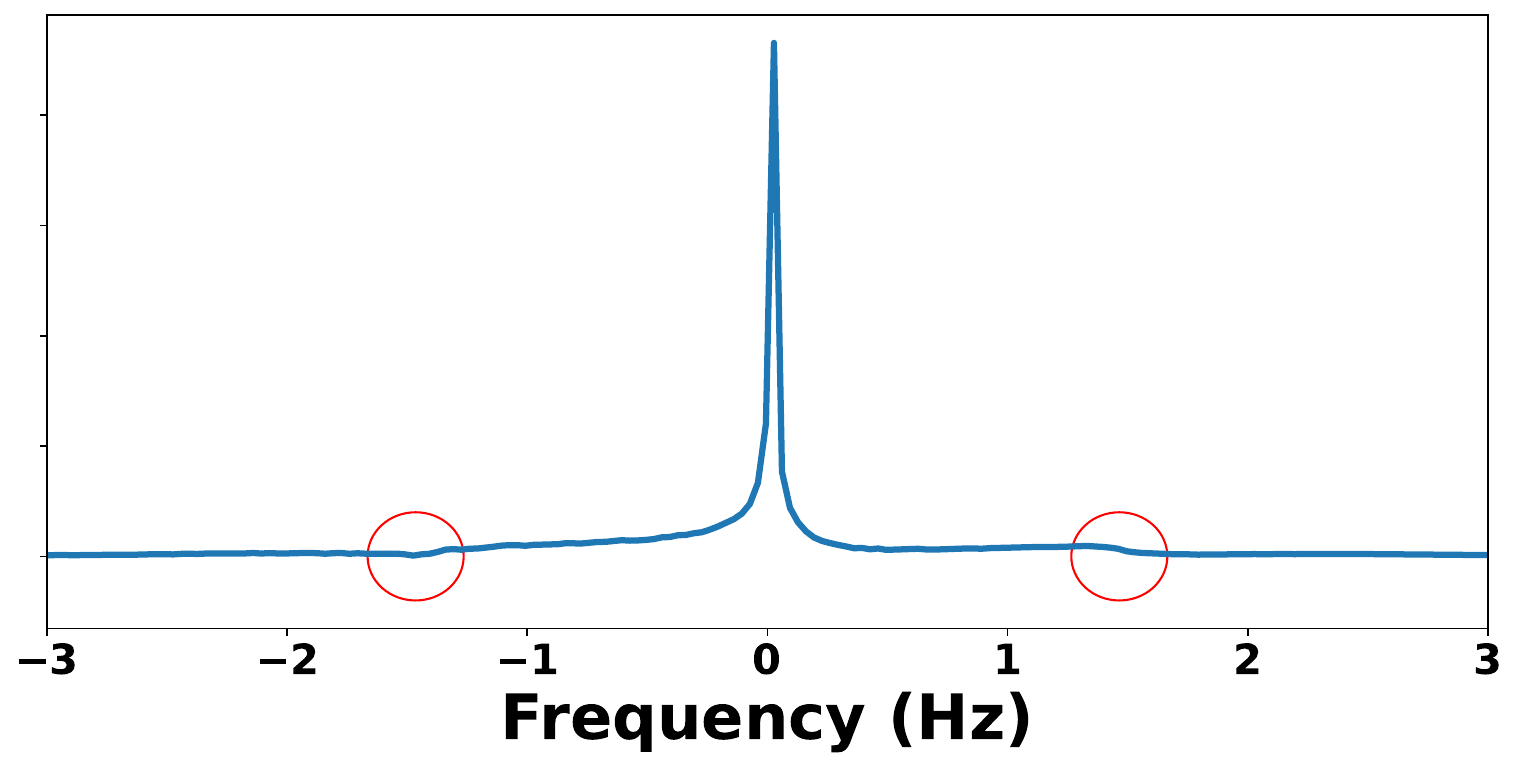}} 
\subfigure[]{\label{fig_SI:B1_measure_nutation_ethaol_sim}\includegraphics[width=.45\textwidth]{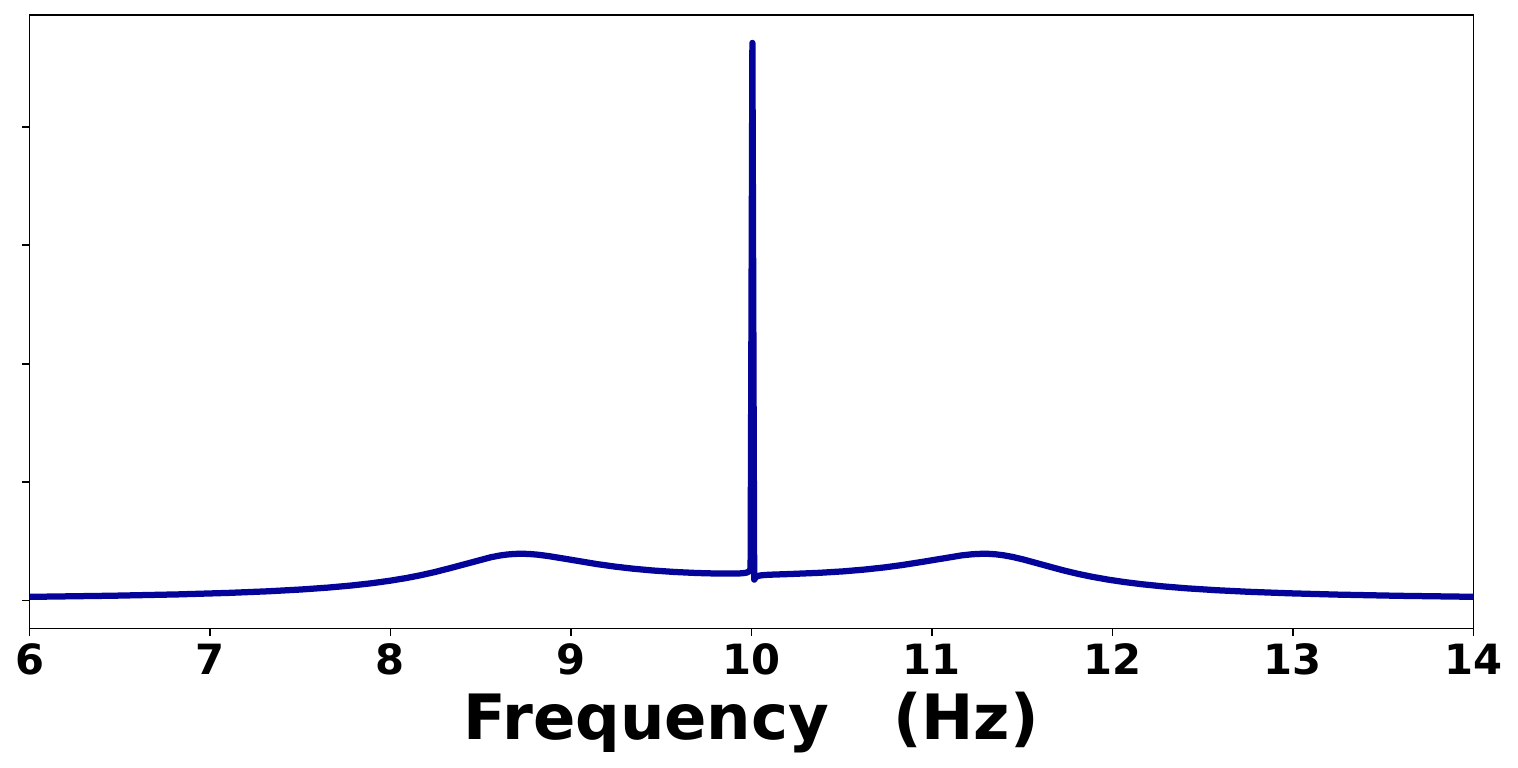}} \\

\caption{\label{fig_SI:B1_measure_nutation} Measurement of the rf leakage from the eFCU by a nutation experiment.
Figure \ref{fig_SI:B1_measure_nutation_methaol} and \ref{fig_SI:B1_measure_nutation_ethaol} are the magnitude spectra corresponding to methanol and ethanol when the LO was set on the CH3 resonance whilst the input of eRFCU was disconnected from the probe during acquisition. 
Two peaks corresponding to the nutation of the spins and caused by the constant rf leakage were respectively observed at $\nu_1=4$ Hz and $\nu_1=1.4$ Hz away from the Larmor frequency of CH3 in methanol and ethanol. 
A simulation was performed using rf leakage amplitude of 4 Hz and 1.4 Hz with frequency 10 Hz with out radiation damping are shown in figures \ref{fig_SI:B1_measure_nutation_methaol_sim} and \ref{fig_SI:B1_measure_nutation_ethaol_sim}. } 
\end{figure}

\newpage
\section*{Figure \ref{fig_SI:B1_measure_nutation} }
\begin{figure}[h!]
\centering
\subfigure[]{\label{fig_SI:Ethanol_lineProfile_OH_REF}\includegraphics[width=.45\textwidth]{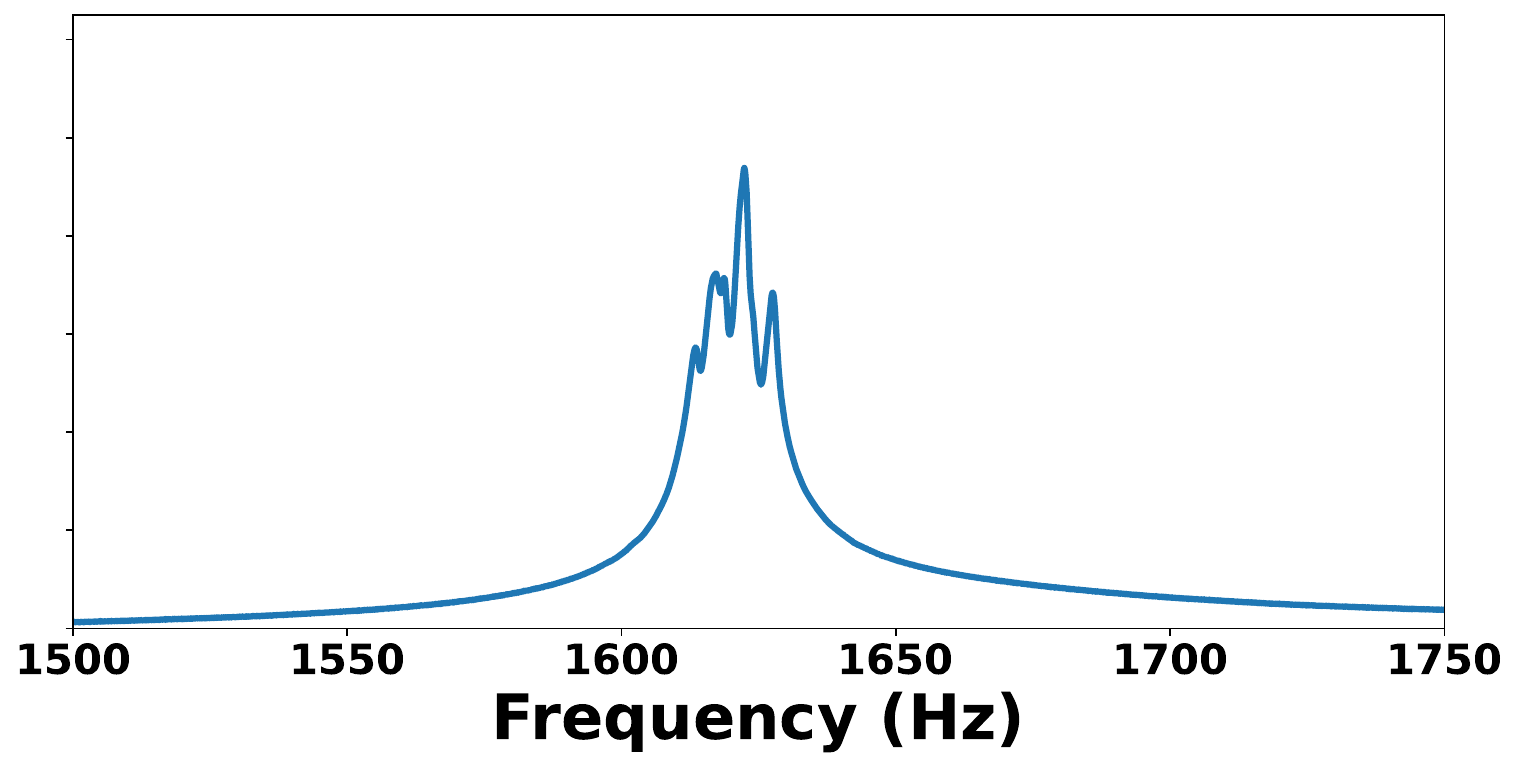}}
\subfigure[]{\label{fig_SI:Ethanol_lineProfile_OH_REF_sim}\includegraphics[width=.45\textwidth]{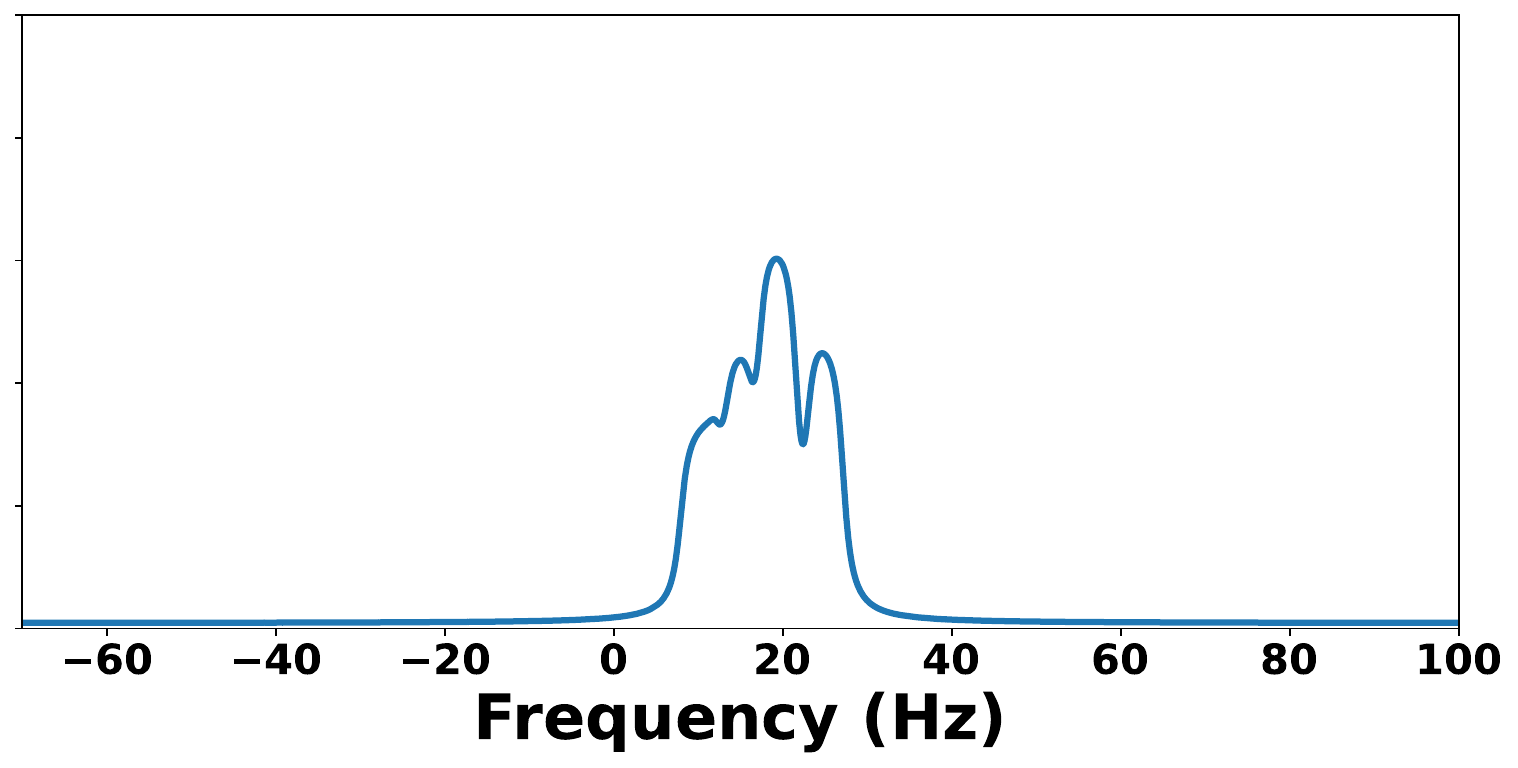}}\\
\subfigure[]{\label{fig_SI:Ethanol_lineProfile_OH_signal}\includegraphics[width=.45\textwidth]{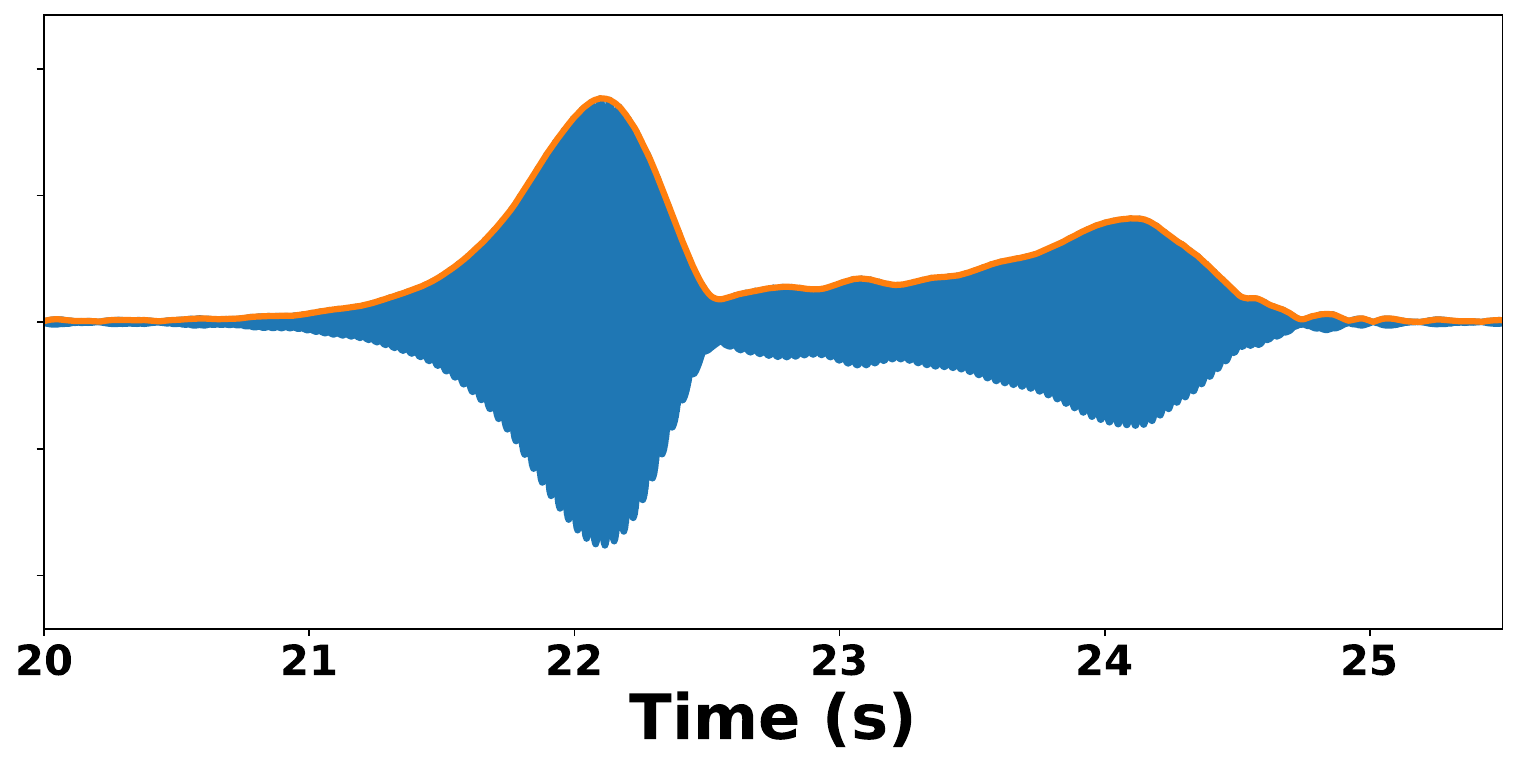}} 
\subfigure[]{\label{fig_SI:Ethanol_lineProfile_OH_signal_sim}\includegraphics[width=.45\textwidth]{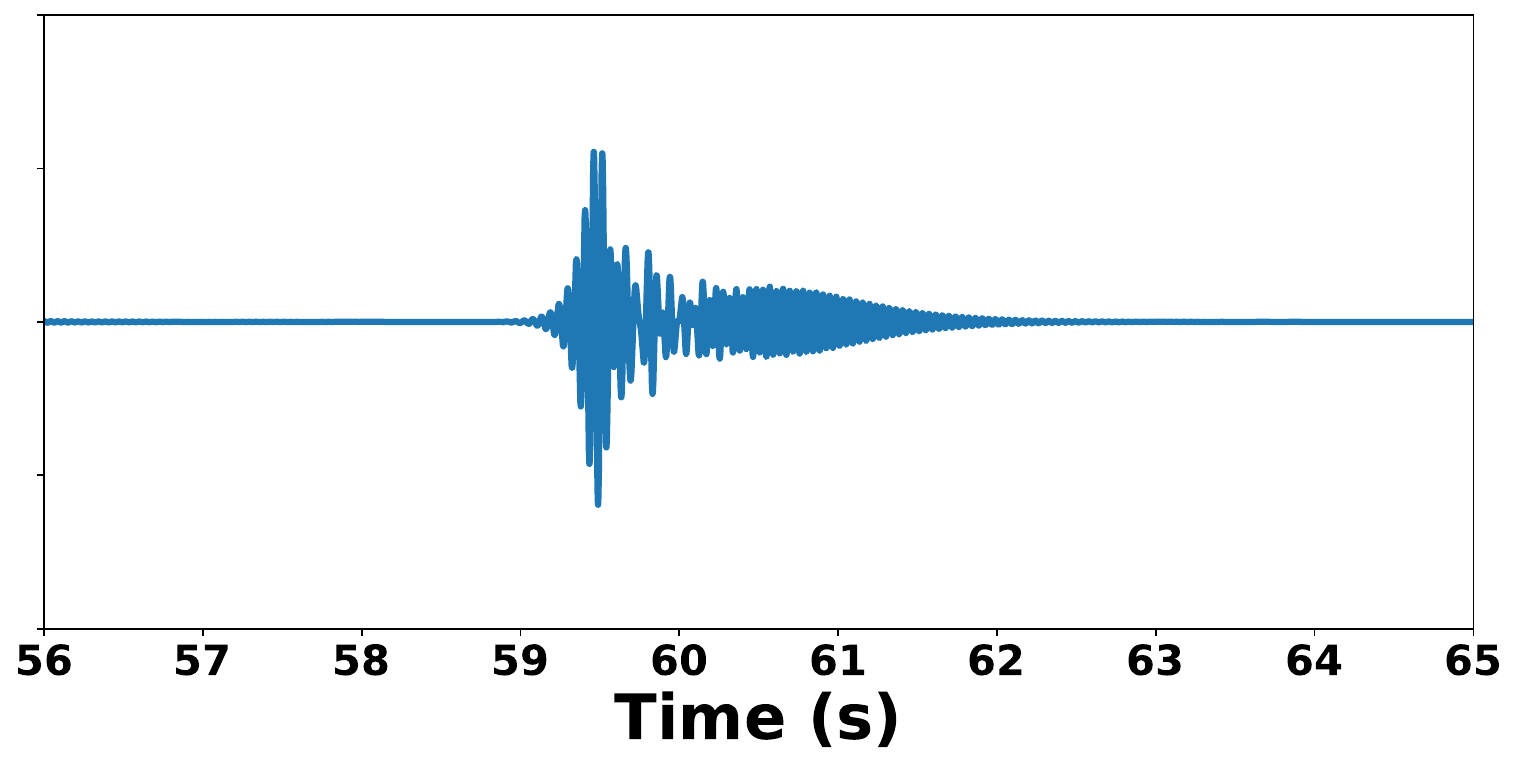}}\\
\subfigure[]
{\label{fig_SI:clustfreq_oh}\includegraphics[width=.45\textwidth]{images/ethanol_OH_proj.pdf}} 
\subfigure[]{\label{fig_SI:Ethanol_lineProfile_OH_signal_sim_oneBurst}\includegraphics[width=.45\textwidth]{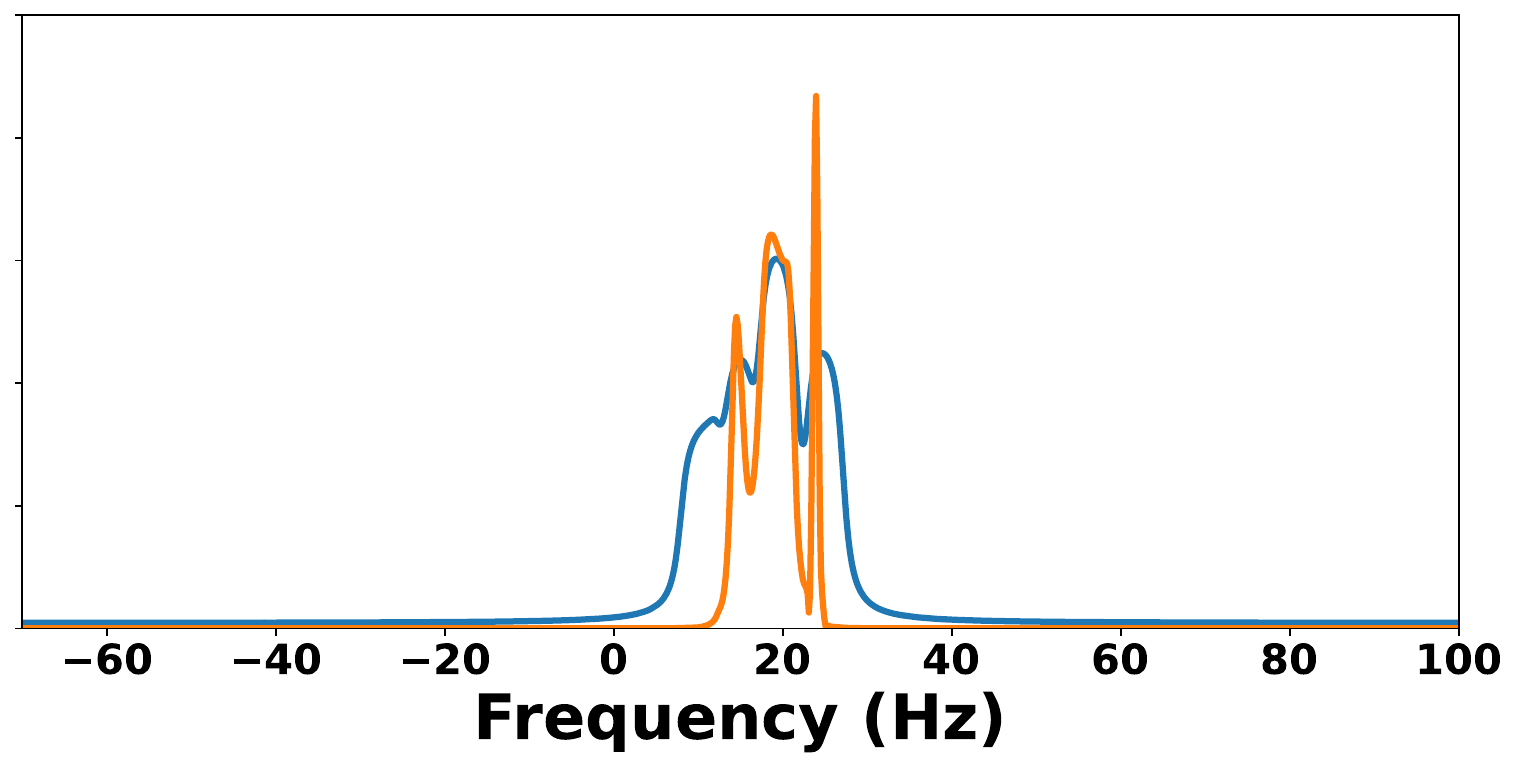}} 
\caption{\label{fig_SI:Ethanol_lineProfile_OH} Maser spectral clustering in ethanol: a toy model simulation versus experiment.
Left column - Figures \ref{fig_SI:Ethanol_lineProfile_OH_REF} corresponds to the spectrum of the OH multiplet acquired after a $90^o$ pulse in the absence of eFBU. In \ref{fig_SI:Ethanol_lineProfile_OH_signal}: single burst from the experimental maser signal shown in figure \ref{fig:multimode_inhomogenity_oh} of the article.
In \ref{fig_SI:clustfreq_oh}, the spectrum in  \ref{fig_SI:Ethanol_lineProfile_OH_REF} and the Fourier transforms of   \ref{fig_SI:Ethanol_lineProfile_OH_signal} show the spectral clustering.
Right column - 
A simulation was performed with five sets of 420 spins separated by 0.01 Hz centered at 10 Hz, 14 Hz, 15.5 Hz, 19.5 Hz and 25 Hz with $m_o$ = 0.002, 0.0014, 0.0014, 0.004, 0.003 to simulate a mock  OH multiplet.
 The associated spectrum and a single burst of the simulated maser signal are shown in \ref{fig_SI:Ethanol_lineProfile_OH_REF_sim}  and \ref{fig_SI:Ethanol_lineProfile_OH_signal_sim}, respectively. 
In \ref{fig_SI:Ethanol_lineProfile_OH_signal_sim_oneBurst} the  spectral clustering from the simulation is shown.} 
\end{figure}

\newpage
\newpage
\section*{Figure \ref{fig_SI:Ethanol_lineProfile_OH}}
\newpage
\section*{Explanation of the methanol spectrum (Fig.\ref{fig:methanol_driaccomb})}
The $\Sha_T$ function, defined as:
\begin{equation}
    \Sha_T = \sum_{n=-\infty}^\infty  \delta(t-n T)
\end{equation}
has the Fourier transform:
\begin{equation}
    \Sha_{1/T} =\frac{1}{T} \sum_{n=-\infty}^\infty  \delta(\nu-\frac{n}{T})
\end{equation}
The induction signal of Fig.\ref{fig:Methanol_limitcycle_signal} can be approximately represented as the sum of two periodic signals of the same period $T$.
Suppose  $f$ and $g$ denote the shape of the signal over one period T and zero elsewhere.
If in addition $g$ is shifted in time, then the signal can be represented as:
\begin{equation}\label{eq_SI:periodicsignal_1}
    S(t) = (f(t)+ g(t+\Delta)) * \Sha_T(t)
\end{equation}
with $\Delta < T$.
The Fourier transform of this signal is thus:
\begin{equation}\label{eq_SI:spec_periodic_1}
    S(f) = \frac{1}{T}\left [ \hat f(\nu) + e^{-2 i \pi \Delta \nu}\hat g(\nu) \right ]\Sha_{1/T}(\nu)
\end{equation}
Now suppose that $g(t)=A f(t)$, $A>0$, and $\Delta=T/2$. The spectrum becomes:
\begin{eqnarray}\label{eq_SI:spec_periodic_2}
\nonumber     S(f) &=& \frac{1}{T}\hat f(\nu)\left [ 1 + A e^{-2 i \pi \nu T/2} \right ]\Sha_{1/T}(\nu)\\
     &=& \frac{1}{T} \sum_{n=-\infty}^\infty   \hat f(n/T)  \left [ 1 + A e^{- i \pi n} \right ]
     \end{eqnarray}
This shows that the coefficient of $\hat f(n/T) $ is equal to $1\pm A$, depending whether $n$ is even or odd. Therefore, the spectral lines are alternatively larger or smaller than the spectrum $f$.
This phenomenon explains the kind of spectrum with alternating intensities observed in Fig.\ref{fig:methanol_driaccomb} of the article.

%
\bibliography{biblio}
\makeatletter\@input{letter_aux.tex}\makeatother